\documentclass[journal]{IEEEtran}
\usepackage{calc,graphicx}
\usepackage{tabu}
\usepackage{epstopdf}
\usepackage[font=footnotesize]{caption}
\usepackage{subcaption}
\usepackage{amsmath}
\usepackage{amsthm}
\newcommand\abs[1]{\left|#1\right|}
\newcommand\norm[1]{\left\lVert#1\right\rVert_{2}}
\newcommand\normf[1]{\left\lVert#1\right\rVert_{2\rightarrow 2}}
\usepackage{algorithm,xpatch}
\usepackage{algcompatible}
\xpatchcmd{\algorithmic}{\setcounter}{\algorithmicfont\setcounter}{}{}
\providecommand{\algorithmicfont}{}
\providecommand{\setalgorithmicfont}[1]{\renewcommand{\algorithmicfont}{#1}}

\usepackage{multirow}
\usepackage{lineno}

\theoremstyle{definition}
\usepackage{amssymb}
\usepackage[T1]{fontenc}

\begin{document}
\setalgorithmicfont{\footnotesize}
\title{Novel Light Weight Compressed Data Aggregation Using Sparse Measurements for IoT Networks}

\author{
\IEEEauthorblockN{Amarlingam M, Pradeep Kumar Mishra, P~Rajalakshmi, Sumohana~S.~Channappayya, and C.~S.~Sastry
}
\thanks{Amarlingam M, Pradeep Kumar Mishra, P Rajalakshmi and Sumohana S. Channappayya are with the Department of Electrical Engineering and C. S. Sastry is with the Department of Mathematics, Indian Institute of Technology Hyderabad, Hyderabad, India (e-mail: ee13p1003@iith.ac.in; ee16mtech11039@iith.ac.in; raji@iith.ac.in; sumohana@iith.ac.in; csastry@iith.ac.in).}

}

\maketitle

\begin{abstract}
Optimal data aggregation aimed at maximizing IoT network lifetime by minimizing constrained on-board resource utilization continues to be a challenging task. The existing data aggregation methods have proven that compressed sensing is promising for data aggregation. However, they compromise either on energy efficiency or recovery fidelity and require complex on-node computations. In this paper, we propose a novel Light Weight Compressed Data Aggregation (LWCDA) algorithm that randomly divides the entire network into non-overlapping clusters for data aggregation. The random non-overlapping clustering offers two important advantages: 1) energy efficiency, as each node has to send its measurement only to its cluster head, 2) highly sparse measurement matrix, which leads to a practically implementable framework with low complexity. We analyze the properties of our measurement matrix using restricted isometry property, the associated coherence and phase transition. Through extensive simulations on practical data, we show that the measurement matrix can reconstruct data with high fidelity. Further, we demonstrate that the LWCDA algorithm reduces transmission cost significantly against baseline approaches, implying thereby the enhancement of the network lifetime.

\end{abstract}

\IEEEpeerreviewmaketitle

\section{Introduction} 
\label{intr}
The sensor nodes used in Internet Of Things (IoT) application deployments such as remote sensing and monitoring are typically inexpensive, untethered and are powered through batteries \cite{IoT}.
However, relaying on battery power limits the lifetime of the nodes. Further, regular recharging or replacement of batteries leads to additional cost and is a laborious task \cite{btry}. Thus, the network lifetime is a critical concern for data aggregation in IoT networks. Wireless transmission consumes significant amount of energy during the data aggregation \cite{Wtrns}. Indeed, reducing the number of packet transmissions and minimizing routing path for data aggregation in the network can improve the network lifetime. Several approaches have been proposed to address this problem \cite{cs01_10}.

\par Compressed Sensing (CS) \cite{cs_intr1} is a signal processing technique that has proven to be very promising for data aggregation \cite{cs_dag_intr2}. CS provides a new perspective for data aggregation in IoT networks enabling the compression and route minimization jointly for energy efficiency over the network \cite{cso}-\cite{cs12}. Most of the CS aided data aggregation techniques use either dense \cite{cso}-\cite{cs12} or sparse random measurements \cite{dsrp3}-\cite{cs13.phi}. These methods have proposed the encoding by utilizing the structural properties of the measurement matrix. In dense random measurements based data aggregation techniques, it is assumed that the individual columns of the measurement matrix are generated at the respective nodes and compute the corresponding measurement \cite{cso}-\cite{cs12}. The sparse random measurements based data aggregation techniques computes the measurements by collecting the data from the interested nodes for each measurement, while assuming that the sparse measurement matrix is stored at each node \cite{dsrp3}-\cite{cs13.phi}. These approaches \cite{cso}-\cite{cs13.phi} aggregate the measurements from all the nodes by minimizing the routing path to reduce the energy consumption in data aggregation.

Most of the existing CS aided data aggregation approaches do not consider the feasibility of hardware implementation \cite{cso}-\cite{cs13.phi}. The bottleneck for hardware implementation of the CS aided data aggregation techniques is in the encoding process at IoT nodes that are severely resource constrained. The size of the measurement matrix depends on sparsity of the sensing data and the number of nodes deployed in the network \cite{cs_dag_intr2}. As IoT nodes are resource constrained devices, for sparse random measurements based data aggregation techniques, storage issues can crop up in large-scale network applications. In case of dense random measurements, the dependency of column size on sensing data sparsity poses multiple constraints in real-time implementation for the applications where data to be sensed has low sparsity \cite{sensor_lit}. In contrast, the measurement matrix content can be combined enroute to the sink instead of generating individual columns or storing the matrix while aggregating the data from the nodes using CS. This class of methods called as routing measurements based data aggregation approaches. Some existing methods in the literature \cite{rout.5ref}-\cite{rndmwlk} have investigated data aggregation using routing measurements. However, these methods compromise either on recovery fidelity (due to low coherence) \cite{rout.5} or energy efficiency (due to higher number of transmissions) \cite{rout.5ref} \cite{rndmwlk}.

  \par Designing a low complexity CS based data aggregation technique that minimizes total energy consumption as well as guarantees the reconstruction is still a challenging problem. To address this problem, in this article, we propose a data aggregation method called ``Light Weight Compressed Data Aggregation (LWCDA)", which is light-weight (low complexity), energy efficient and provides good recovery fidelity. In contrast to some existing approaches \cite{rout.5ref}-\cite{rndmwlk}, we utilize clustering for data aggregation which is proven to be promising for energy efficient routing \cite{cs13.phi} \cite{recent_ch}. In addition, the aggregated data from cluster heads is collected using a minimum spanning tree to minimize energy consumption. In the proposed algorithm, each node measures a data sample followed by generating a random value from a Bernoulli distribution for computing the measurement. The cluster heads receive the measurements from their descendants, process them to compute the final measurement before transporting it to the sink. We find that the measurement matrix constructed from our algorithm is highly sparse and possesses properties to guarantee the recovery of data such as high incoherence, good recovery region and satisfy the Restricted Isometry Property (RIP) when combined with some popular bases.
 
The contributions of this article are summarized as follows:
 \begin{enumerate}
 \item Low complexity CS aided data aggregation technique that constructs a sparse measurement matrix from the network. 
 \item Performance evaluation of the measurement matrix with respect to RIP, coherence and phase transition. 
 \item Comparative analysis of the algorithm in terms of reconstruction error and transmission cost using real data sets. 
 \item A practical implementation using IITH Motes \cite{mote} to demonstrate hardware feasibility of the proposed LWCDA algorithm.
 \end{enumerate}
 \par The paper is organized as follows: Section II explains the basics of compressed sensing in IoT networks. Section III describes the proposed LWCDA data aggregation method. Section IV evaluates the RIP and coherence of the proposed measurement matrix and presents the phase transition analysis. Simulation results of LWCDA method are described in Section V and Section VI describes the hardware implementation performed. Section VII concludes the paper. 
 
\section{Compressed Sensing For IoT Networks}
\label{cs}
\subsection{Compressed Sensing: A review}
For a given $N$ dimensional signal (hereafter data and signal are used interchangeably) that can be sparsely represented using a basis, CS promises to deliver a full recovery of the signal with high probability from far fewer samples \cite{csbk}. Let $X=[x_{1}, x_{2}, x_{3},\ldots, x_{N}]^{T}\in \mathbb{R}^{N}$ be sparsely represented in a basis (e.g., Discrete Cosine Transform (DCT), Discrete Fourier Transform (DFT), Discrete Wavelets Transform (DWT), etc.) $\Psi=[\psi_{1}, \psi_{2}, \ldots, \psi_{N}] \in \mathbb{R}^{N \times N}$ with $k$ large coefficients ($k$-sparse), where $k\ll N$, i.e., $X=\Psi\theta$, $\theta= [\psi_{1}^{T} X, \psi_{2}^{T} X, \dots, \psi_{N}^{T} X] \in \mathbb{R}^{N}$ and $\|{\theta}\|_{0} \leq k$. The CS theory computes the compressed $M$-dimensional vector,
\begin{equation}
Y= \Phi X,
\label{eq1}
\end{equation}
where $Y \in \mathbb{R}^{M}$ is the measurement vector and $M$ is the number of measurements and $M<N$ which influences reconstruction of the signal. It has been shown that the number of random measurements required for successful reconstruction of a $k$-sparse signal is $M= \mathcal{O} ( k$ $log$ $N)$ \cite{mk}. The matrix $\Phi=[ \varphi_{1}^{T},\varphi_{2}^{T}, \ldots, \varphi_{M}^{T} ]^{T} \in \mathbb{R}^{M \times N}$ is called the measurement matrix. The problem here is to reconstruct $X$ from $Y$, which is under-determined and can have infinitely many solutions. CS theory shows that the problem of recovering $X$ from its linear measurements can be posed as a $l_0$-minimization problem as shown in \eqref{eq3} and it is computationally intractable. A family of greedy algorithms have been proposed in \cite{greedy} and \cite{OMP} to solve the $l_0$-minimization problem.

\begin{equation}
 \min_{\theta} \|\theta\|_{0}  \text{ subject to} \quad \Phi\Psi\theta = Y.
 \label{eq3}
\end{equation}

The most prevalent decoding technique to solve the problem in \eqref{eq3} is $l_{1}$-minimization, which is a convex optimization problem \cite{l0_l1a} and hence, computationally tractable \cite{cvx},
\begin{equation}
 \min_{\theta} \|\theta \|_{1}  \text{ subject to} \quad \Phi\Psi \theta=Y.
\label{eq2}
\end{equation}

From the solution $\theta$ obtained using $l_{0}$ or $l_{1}$-minimization, $X$ can be reconstructed as,
\begin{equation}
\widehat{X} = \Psi \theta.
\label{eq4}
\end{equation} 
 
 The CS matrix $A=\Phi\Psi$ plays a crucial role in the recovery of the $N$ dimensional original signal $X$. In \cite{RIP1}, it is shown that the CS matrix $A$ should satisfy the property known as RIP for successful recovery of $X$ using $l_{1}$ minimization. A matrix $A\in \mathbb{R}^{M \times N}$ is said to satisfy the RIP of order $k$ with constant $\delta_{k} \in (0,1)$ if
\begin{equation}
(1-\delta_{k})\norm{u}^{2}\leq \norm{Au}^{2}\leq (1+\delta_{k})\norm{u}^{2}, \forall u \in \Sigma_{k},
\label{rip}
\end{equation}
where $u$ is a $k$-sparse vector and $\Sigma_{k}$ is set of all $k$-sparse vectors.

\par On the other hand, if $X$ can be sparsely represented in $\Psi$ domain, then to achieve successful recovery, the theory of CS requires low mutual coherence between the columns of the CS matrix $A=\Phi\Psi$. The mutual coherence of the CS matrix can be defined as
\begin{equation}
\mu (A) =\max \limits_{1\leq p \neq q \leq N}\abs{\langle a_{p},a_{q}\rangle},
\label{cohr}
\end{equation}  
where $a_{p}$ and $a_{q}$ are normalized columns of $A$. 

\subsection{Related Works}
\label{rwrk}
In this section we discuss the contributions of the relevant literature. Most of the CS aided data aggregation techniques can be classified into three classes, dense random measurements \cite{cso}-\cite{cs12}, sparse random measurements \cite{dsrp3}-\cite{cs13.phi} and routing measurements \cite{rout.5ref}-\cite{rndmwlk} based data aggregation methods. 

Dense random measurements based methods \cite{cso}-\cite{cs12} achieve CS aided data aggregation by considering individual column generation of the measurement matrix at node level using pseudo-random sequences. These methods aggregate the measurements from all the nodes by minimizing routing path to achieve energy efficiency. The size of the measurement matrix depends on the number of nodes and sparsity of the data. IoT nodes are constrained devices possessing minimal on-board resources (in terms of physical memory, processing capability, internal memory, energy). Therefore, generating individual columns of the measurement matrix at a node in case of a large-scale network application where sensing data sparsity is low is computationally intensive and poses multiple constraints in real-time implementation. 

Wang $et$ $al. $ \cite{dsrp3} showed that sparse random measurements (projections) reduce communication cost per sensor node for data aggregation. In \cite{cs1}-\cite{cs13.phi}, data aggregation techniques have been proposed to achieve energy efficiency for IoT networks by using the sparse random measurements \cite{dsrp3}. These algorithms find the optimal route to collect data from the interested nodes for each measurement, while assuming that the sparse measurement matrix is stored at each node. Since the measurement matrix depends on the network size, storage issues can crop-up for large-scale networks. In other words, commercially available nodes that have minimal on-board resources may not be able to support the storage large measurement matrices.

In contrast to dense and sparse random measurements, the routing measurements based data aggregation methods aggregate the measured data from the nodes by computing measurements on the fly enroute to the sink \cite{rout.5ref}-\cite{rndmwlk}. In \cite{rout.5ref}, the routing paths are iteratively built through a greedy choice to minimize the coherence of the CS matrix and energy required for data aggregation. However, building of routing paths in an iterative manner is computationally intensive and requires more transmissions rendering the process highly energy inefficient. In \cite{rout.5}, the algorithm picks up a portion of the nodes randomly from the network to generate measurements by utilizing shortest path routing. However, such an approach does not achieve good performance with respect to coherence. In \cite{rndmwlk}, the authors showed that data aggregation from fixed length random walks starting at randomly located nodes can reconstruct the data using CS. However, recovery performance of the method depends on the length of the random walks. An increase in the length of the walk increases the number of transmissions which in turns data aggregation to be energy inefficient.
 
\subsection{Problem Statement}
 As discussed above, CS based data aggregation algorithms proposed in the literature do not address the aspects of low complexity and energy efficiency jointly. The approaches proposed based on dense random measurements \cite{cso}-\cite{cs12} as well as sparse random measurements \cite{dsrp3}-\cite{cs13.phi} are energy efficient but not real-time implementable. On the other hand, the approaches proposed in \cite{rout.5ref}-\cite{rout.5} are light weight, however they are either energy inefficient or do not achieve good performance in terms of coherence and recovery. These limitations provide the motivation for this work. Specifically, the problem is to design a low complexity (real-time) CS aided data aggregation method that is energy efficient and can guarantee a successful recovery of the data for IoT networks.

\section{Proposed Data Aggregation Protocol}
In this section we first present the network model that will be used in our analysis and next describe the proposed data aggregation protocol which forms the light weight measurement matrix.
\subsection{Network Model}
Consider an IoT network with $N$ nodes deployed in a rectangular area (an example network with grid-wise deployment of $N=100$ nodes is shown in Fig. \ref{phi1}). The network can be represented by a graph $G(V,E)$, where $V$ is the set of vertices or nodes and $E$ represents the set of edges or links between the nodes. The sink node $S$ is the node that collects data from all the other nodes in the network. We assume that all the nodes are loosely time synchronized and have homogeneous transmission coverage. Unit disc coverage model is considered for all the nodes. We consider the communication range of the nodes to be $D=~\sqrt{\frac{5}{N}}*a$ \cite{rout.5}. 
Here, $a$ is the length of the maximum side of the considered area and $N$ is the number of nodes. Data aggregation proceeds in cycles (rounds) and each node generates one sample per cycle. For example, the $i^{th}$ node acquires data sample $x_i$ in each cycle and $N$ samples $X=[x_{1}, x_{2}, x_{3},\ldots, x_{N}]^{T} \in \mathbb{R}^{N}$ will be acquired from all the nodes per cycle. We also assume that there is no packet loss in data aggregation. We consider both grid \cite{grid_dplmnt1} \cite{grid_dplmnt2} and random deployment \cite{rndm_dplmnt} scenarios for analysis in further sections as these network deployments have their own significance in different application scenarios.
 
\subsection{Proposed Data Aggregation Protocol}
As described above, $X \in \mathbb{R}^{N}$ is a signal of length $N$ that contains measurements from $N$ nodes in the network. To aggregate data from all the nodes, $M$ nodes are randomly picked such that each node is a Cluster Head (CH) with a probability $P_{CH}=\frac{M}{N}$. The remaining ($N-M$) leaf nodes connect to their respective nearest CH through the shortest path. Accordingly, the whole network gets divided into $M$ non-overlapping clusters to aggregate sensors data. The $M$ clusters $\{c_{1},c_{2}, \ldots,c_{M}\}$ can contain distinct $\{n_{1},n_{2}, \ldots, n_{M}\}$ number of nodes. Every node in the cluster measures its data sample $x_{i}$ (e.g., temperature, humidity, light intensity, etc.) and multiplies it with a random value $\alpha_{i}$ generated from a Bernoulli distribution with a success probability of $0.5$. In other words, the $i^{th}$ node performs $\alpha_{i}x_{i}$, where $\alpha_{i}$ is randomly drawn from the set $\{-1,1\}$ with a Bernoulli distribution and $i \in [1,N]$. Each leaf node sends the measurement $\alpha_{i}x_{i}$ to its CH. The CH adds the received measurements from the leaf nodes including its own measurement. The final measurement at $j^{th}$ CH, $y_{j}=\sum_{i\in c_{j}}\alpha_{i}x_{i}$ is the linear combination of $\alpha_{i}$ and $x_{i}$, where the nodes belonging to the cluster take non-zero values i.e., $\{\alpha_{i}\neq0, x_{i}\neq0 \}\in c_{j}$ and the nodes that do not belong to the cluster can be assumed to be zeros i.e., $\{\alpha_{i}=0,x_{i}=0\} \notin c_{j}$. The CHs deliver the computed measurements to the sink node through the Minimum Spanning Tree (MST). Dijkstra's and Kruskal algorithms can be used to create MST of CHs along with the sink node. The CHs follow the pack and forward method \cite{cs01} that provides the feasibility to encapsulate the current measurement of a CH with the relaying packet from descendant CHs along the MST towards the sink.  
 
From the CS formalism in Section II, each cluster can be considered as a row of the measurement matrix $\Phi$ and each node in the network corresponds to a column of $\Phi$. In other words, $M$ randomly formed clusters and the nodes in each cluster correspond to rows and respective columns of $\Phi$. The $j^{th}$ cluster $c_{j}$ forms the $j^{th}$ row of $\Phi$, i.e., $\varphi_{j}$. The support vector of $\varphi_{j}$ is $\Delta_{j}=\{i: i\in[1,N], i\in c_{j}\}$, $\varphi_{j_{\Delta_{j}}} =\{\alpha_{i} : i \in \Delta_{j}\}$ and $\varphi_{j_{\Delta_{j}^{c}}}=0$. In other words, the $j^{th}$ row of $\Phi$ at respective columns of nodes that are connected as a cluster $i \in c_{j}$ will be assigned values from the set $\{-1,+1\}$ with a Bernoulli distribution. The remaining entries in the row will be zeros.

More concretely, $\Phi \in \mathbb{R}^{M \times N}$, $\Phi=[ \varphi_{1}^{T},\varphi_{2}^{T}, \ldots, \varphi_{M}^{T} ]^{T}$ contains elements in each row

\[
  \varphi_{ji}= 
  \begin{cases} 
      -1 \text{ or } +1 & \text{if } i\in c_{j} \\
       0               &\text{otherwise }. 
   \end{cases}
\]

Packets received at the sink node from the MST contain elements of the measurement vector $Y=[y_{1}, y_{2}, \ldots, y_{M}]^{T} \in \mathbb{R}^{M}$ which are linear combinations of the measured data and the random values of nodes, i.e.,

\begin{equation}
Y=
\begin{pmatrix}
  y_{1}\\y_{2}\\\vdots\\y_{M}
\end{pmatrix}
=
\begin{pmatrix}
  \varphi_{1} \\ \varphi_{2} \\ \vdots \\ \varphi_{M}
\end{pmatrix}
\begin{pmatrix}
  X
\end{pmatrix}
=
  \Phi X,  
\label{csm}
\end{equation}

where $X=[x_{1},x_{2}, \ldots, x_{N}]^{T}$, $X \in \mathbb{R}^{N}$, $\varphi_{m} \in \mathbb{R}^{N}$, $y_{m} \in \mathbb{R}$ where $m \in [1,M]$.
 
\begin{figure*}[ht]
\centering
\includegraphics[scale=0.36]{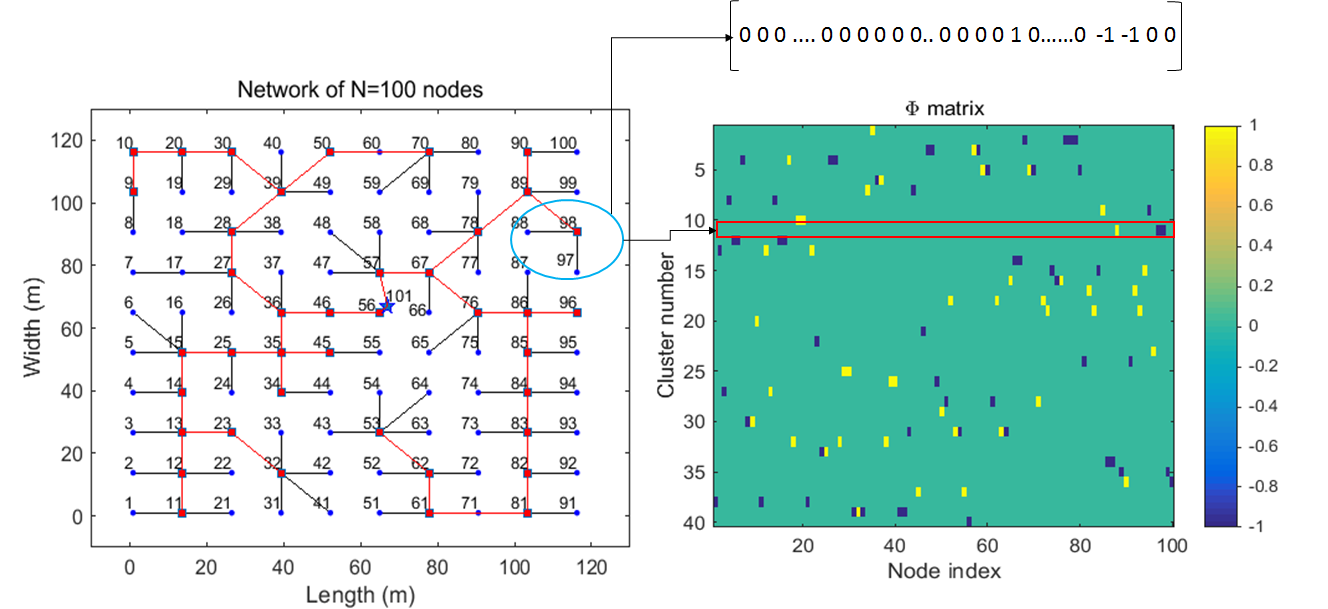}
\caption{The procedure of measurement matrix designing from a network with $N=100$ nodes and $M = 40$. The sink $S=101$ is represented by a star, square boxes represent the CHs and remaining nodes are leaf nodes.}
\label{phi1}
\end{figure*}

To gain insight into the described LWCDA, we consider a network of grid-wise deployed 100 nodes with a sink node (S = 101), which is placed at the center of the network as shown in Fig.~\ref{phi1}. Fig.~\ref{phi1} shows the measurement matrix formation from the network and the sink node. Consider the example node of $98$ from Fig.~\ref{phi1}, which is a CH and has two descendant nodes $88, 97$. The measurement matrix $\Phi$ contains a row which represents the cluster with the nodes $98, 88, 97$ and contains non-zero values from the set $\{-1,+1\}$ that are drawn from a Bernoulli distribution with a success probability of $0.5$ at respective columns, as shown in Fig. \ref{phi1}.

To recover the original signal $X \in \mathbb{R^{N}}$ from the measurement vector $Y \in \mathbb{R}^{M}$, the sink node needs to have the knowledge of $\Phi$. The information of $\Phi$ can be shared with the sink by maintaining synchronized seeds and pseudo-random number generators between the nodes and the sink \cite{cso}. Practically, to share pseudo-random number seeds, each node has to send its seed to the sink \cite{rndmwlk} and this requires a large number of transmissions for large-scale networks. Another approach is to transmit information of $\Phi$ to the sink along with the measured data if the message overhead is negligible \cite{rout.5}. In our data aggregation algorithm, each node needs to send or share the information of $\alpha =-1$ or $+1$ with the sink, which can take a maximum of one octet of packet payload. We consider the case that every $j^{th}$ CH sends individual $\alpha$ values of the nodes that belong to that cluster $c_{j}$ and their indices $\Delta_{j}$ along with the measurement $\sum_{i\in c_{j}}\alpha_{i}x_{i}$ to the sink in the first cycle of data collection. The system of linear equations in \eqref{csm} ($M<N$) is under-determined and will give infinitely many solutions while recovering $X$ from $Y$. The sink node reconstructs full dimension $\widehat{X} \in \mathbb{R}^{N}$ from the received measurement vector $Y \in \mathbb{R}^{M}$ by solving either of the optimization problems discussed in Section \ref{cs}.

\subsubsection{Node-level Complexity for Encoding}
\label{cmplxty}
The node-level complexity of measuring the data is computed in terms of generating or storing the number of random values. The proposed data aggregation algorithm constructs $\Phi$ on the fly while data is being aggregated from the nodes. Note that each node is required to generate only a random value $-1$ or $+1$ from a Bernoulli distribution as discussed above. The node level complexity of our method in terms of generating or storing number of random values is $\Theta{(1)}$ which is independent of sensing data sparsity and network size. The $\Theta(\hspace{0.1cm})$ refers the formal notation for stating the exact bound on growth of resource needs (computation and storage) of an algorithm. 
Baseline data aggregation approaches \cite{cso}-\cite{cs12} which utilize dense random measurements require the generation of the respective columns at each node which is the size of $\Theta{(M)}$ units. In case of sparse random measurements based data aggregation methods \cite{cs1}-\cite{cs13.phi}, every node in the network stores the complete $\Phi$. The size of the required storage is $\Theta{(MN)}$. Some of the methods which use sparse random measurements such as \cite{dsrp3} generate the respective row of $\Phi$ at every node and the row size is $\Theta{(N)}$ units. The values of $M$ and $N$ are proportionally related and depend on sensing data sparsity and the network size. This dependency poses multiple constraints on the real-time implementation of the large-scale network applications where the data to be sensed has low sparsity. The proposed approach is lightweight as it completely eliminates the burden of generating a specific column or storing the entire $\Phi$ at the node to perform data aggregation in the network. Consequently, the proposed method can be implemented in commercially available low end IoT nodes.

The measurement matrix $\Phi$ should satisfy certain properties as discussed in Section \ref{cs} for it to allow data recovery. In the following section we evaluate the properties of the $\Phi$ and demonstrate how it can guarantee the reconstruction.

\section{Measurement Matrix Analysis}
To analyze the proposed measurement matrix $\Phi$, we rely on RIP, coherence and Phase Transition (PT) \cite{Pt} analyses. We considered both grid and random deployments scenarios as both deployments have their own significance for different application scenario \cite{grid_dplmnt1}-\cite{rndm_dplmnt}. We considered DCT, DFT, DWT, Laplacian and Diffusion Wavelet (DiWT) bases ($\Psi$) for the analyses. The DCT, DFT and DWT bases ($\Psi$) can sparsify data from regular (grid-wise) IoT deployments \cite{cso}, \cite{rndmwlk}. In case of randomly deployed networks, the Laplacian \cite{rndmwlk} and Diffusion wavelet (DiWT) \cite{diwt_basis} can accommodate irregularity and provide a sparse representation of the data. 
\subsection{Numerical Experiments: RIP Analysis}
As discussed in Section \ref{cs}, RIP is a standard tool to analyze near-orthonormal performance of a CS matrix while operating with sparse input vectors. This property measures the performance of a compressed sensing matrix in terms of the Restricted Isometry Constant (RIC) $\delta_{k}$. As a result, $\delta_{k}$ can be used to evaluate the ability to recover a sparse signal from the measurement vector. From the definition of RIP of a matrix $A \in \mathbb{R}^{M\times N}$, for $k$-sparse vectors with a constant $\delta_{k}$, \eqref{rip} can be rewritten as, 
\begin{equation}
\delta_{k}= \max \limits_{T \subset [N], |T| \leq k}\normf{A_{T}^{*}A_{T}-Id}, 
\label{rip1}
\end{equation}
where $Id \in \mathbb{R}^{|T|\times |T|}$ is an identity matrix and $T$ is the support set of $k$-sparse vector \cite{csbk}. 

\par For any matrix $A$ that satisfies RIP with a RIC of $\delta_{k}$, the following condition holds:
\begin{equation}
(1-\delta_{k})\leq \lambda_{min} (A_{T}^{*}A_{T}) \leq \lambda_{max}(A_{T}^{*}A_{T}) \leq (1+\delta_{k}),
\label{rip2}
\end{equation}
where $\lambda_{min}$ and $\lambda_{max}$ are the minimum and maximum eigenvalues of the symmetric matrix $A_{T}^{*}A_{T}$ respectively. 

\begin{figure*}[pt]
\centering
        
         \begin{subfigure}[b]{0.4\textwidth}
         	\includegraphics[width=\linewidth]{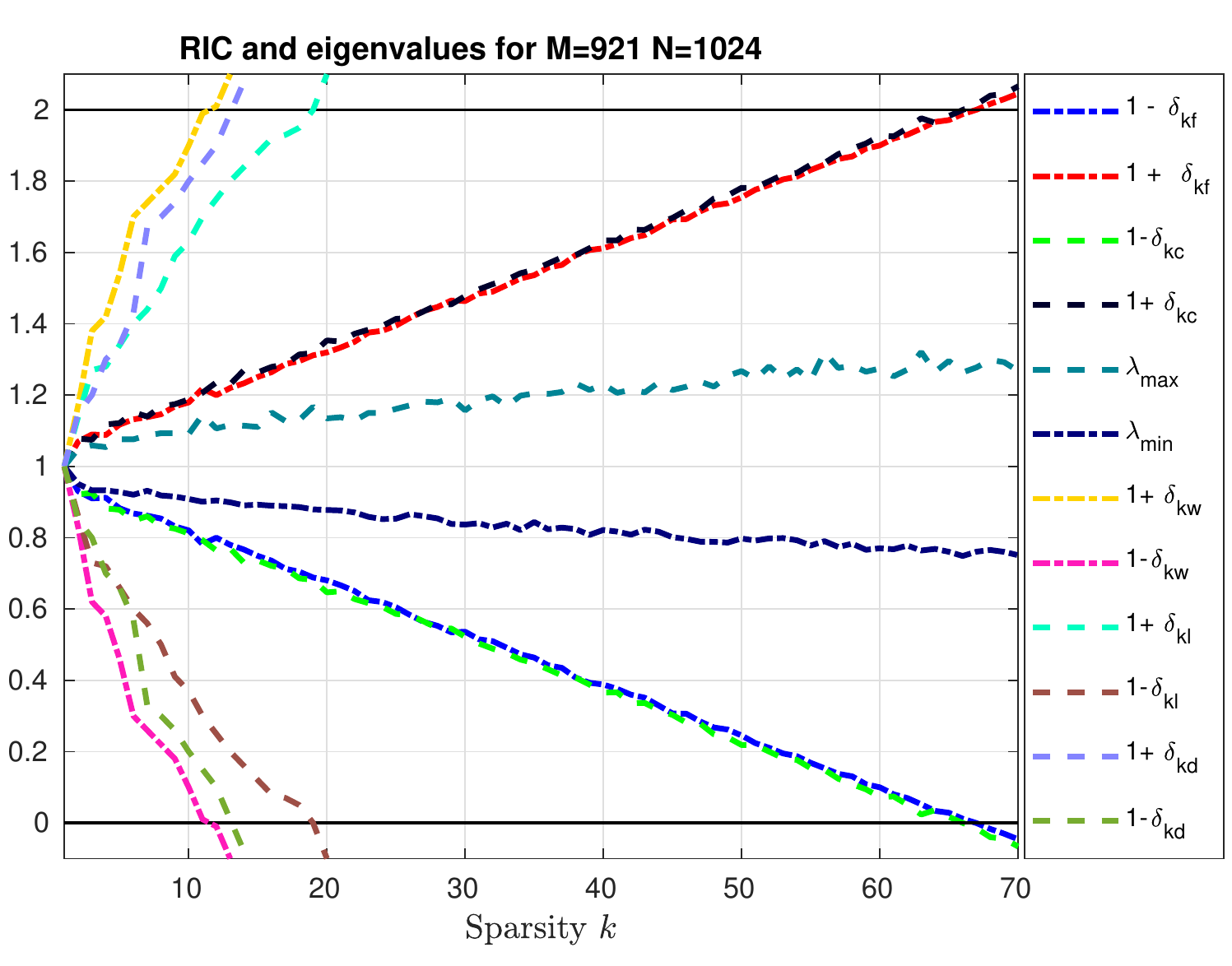}
         	\caption{$\Gamma=10\%$.}
         	\label{1}
         \end{subfigure}%
         \begin{subfigure}[b]{0.4\textwidth}
         	
         	\includegraphics[width=\linewidth]{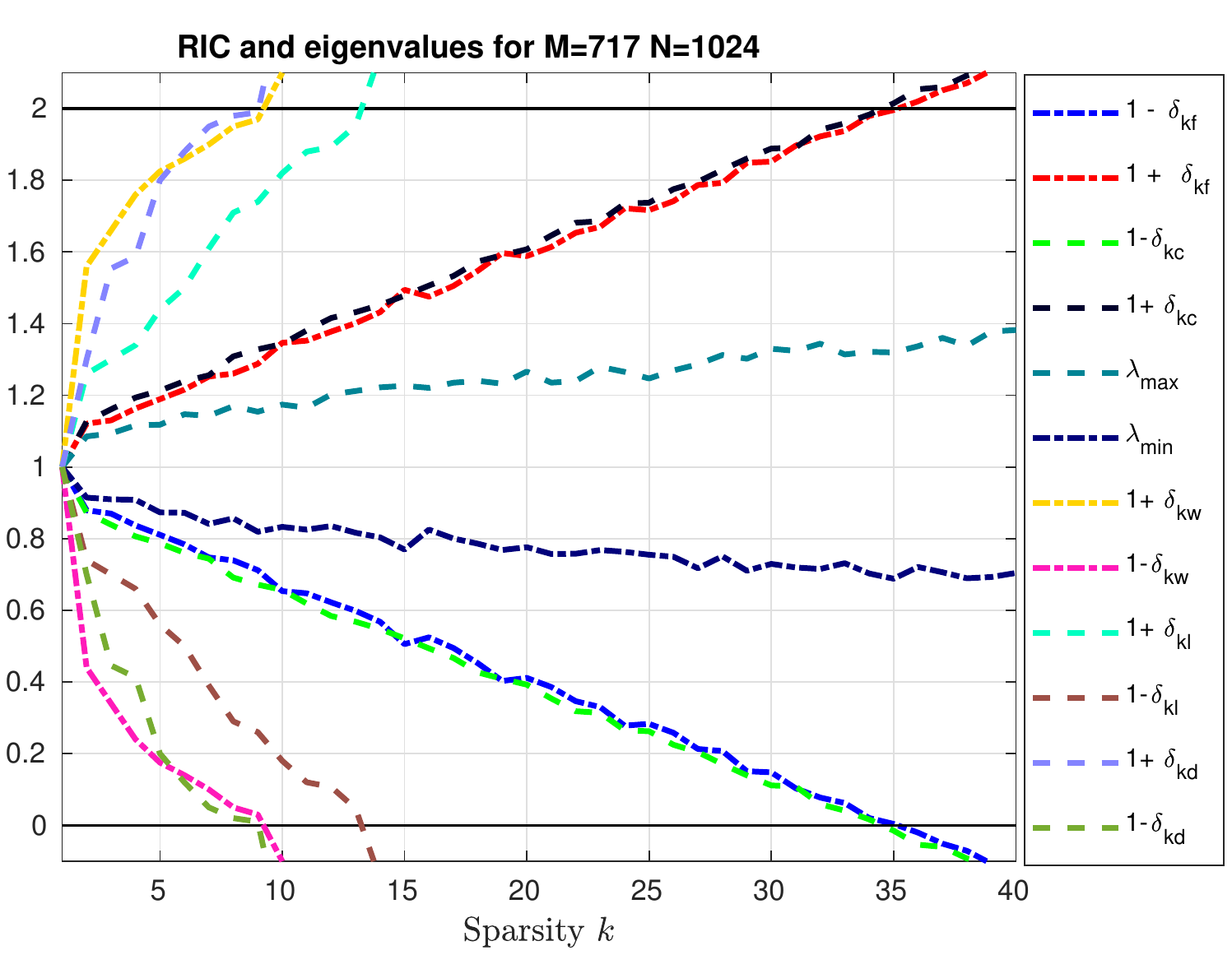}
         	\caption{$\Gamma=30\%$.}
         	\label{2}
         \end{subfigure}
         \begin{subfigure}[b]{0.4\textwidth}
         	
         	\includegraphics[width=\linewidth]{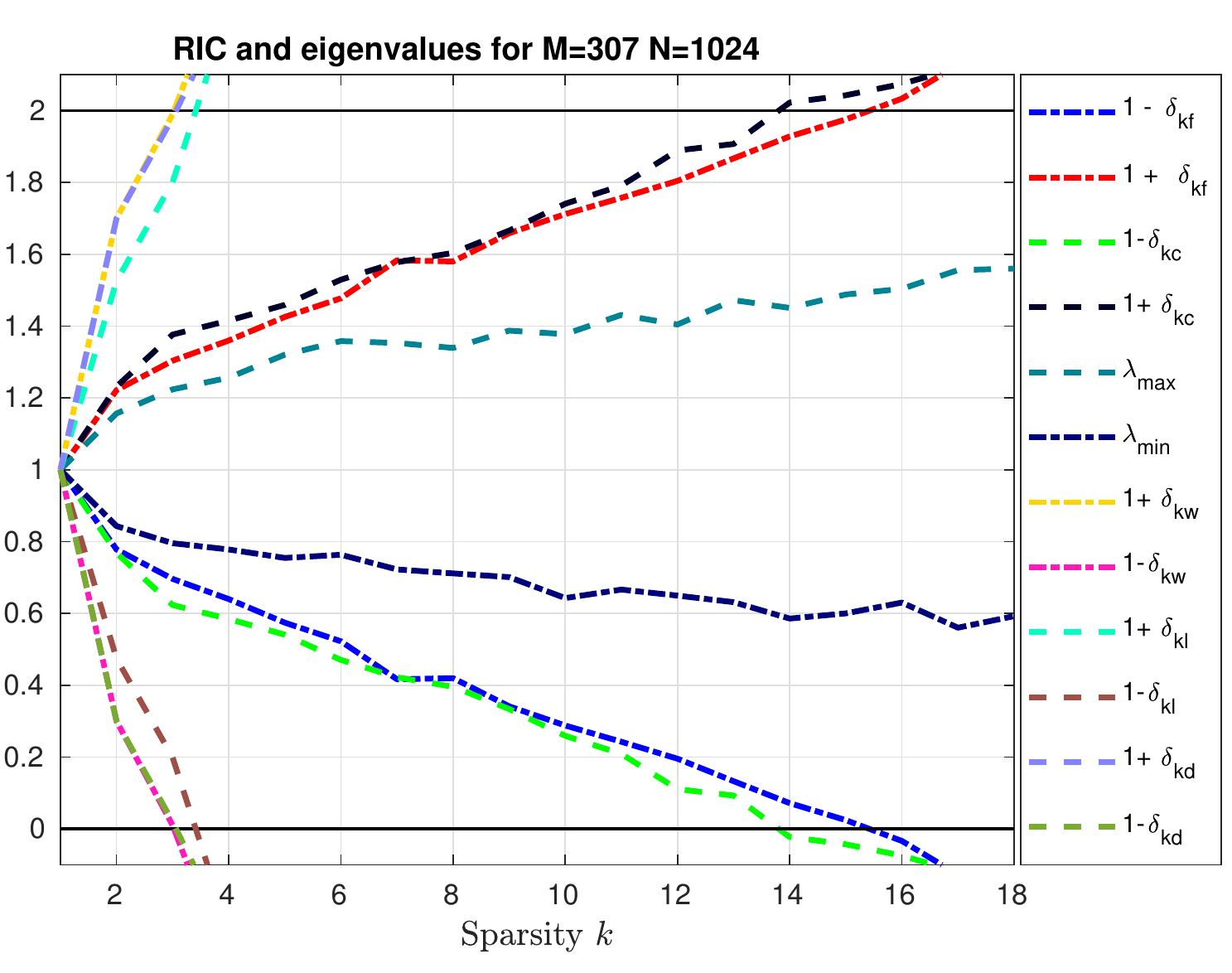}
         	\caption{$\Gamma=70\%$.}
         	\label{4}
         \end{subfigure}%
         \begin{subfigure}[b]{0.4\textwidth}
         	
         	\includegraphics[width=\linewidth]{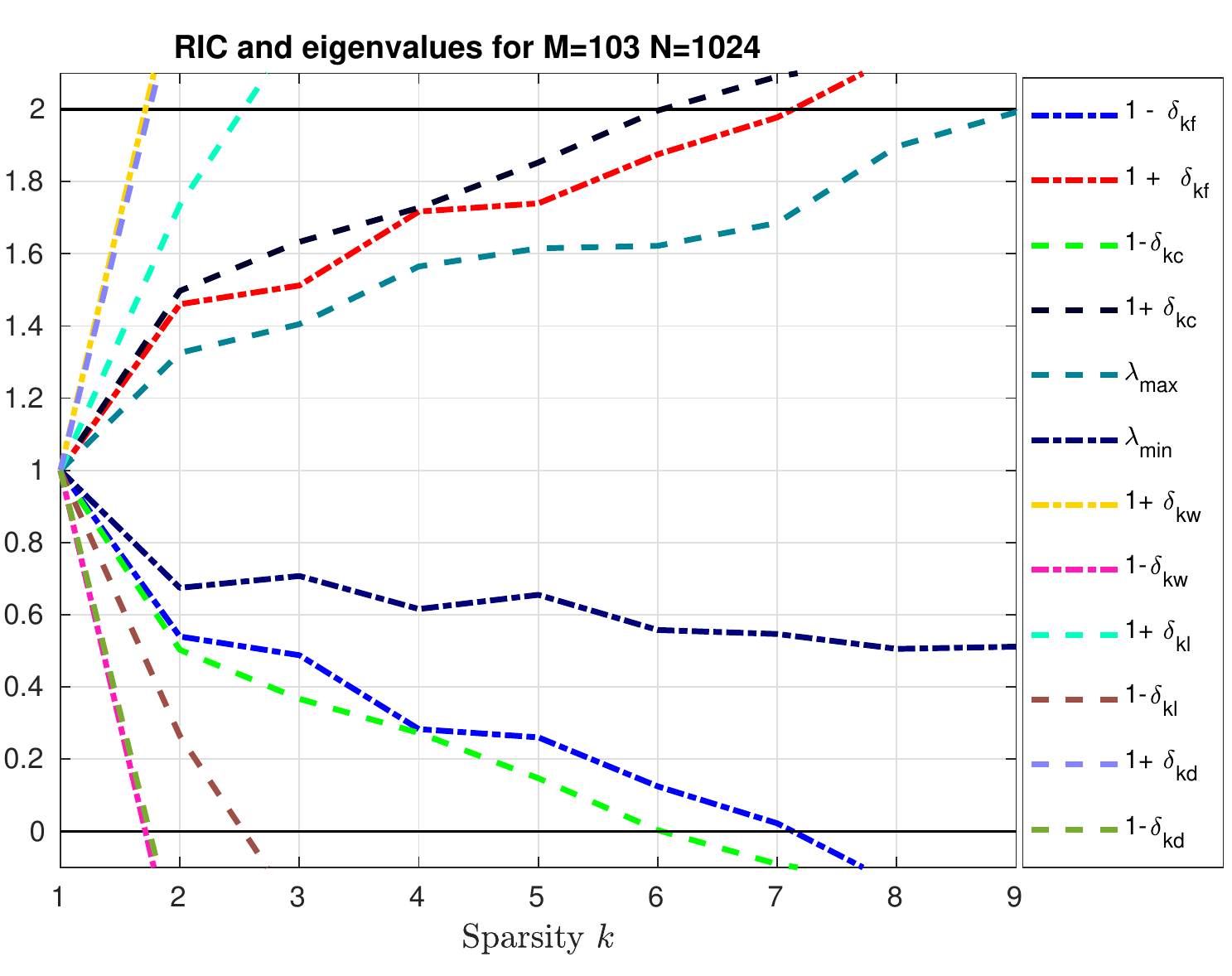}
         	\caption{$\Gamma=90\%$.}
         	\label{5}
         \end{subfigure}
       \caption{RIC $\delta_{k}$ and eigenvalue bounds ($\lambda_{min}$, $\lambda_{max}$) for the proposed CS matrix $A=\Phi\Psi$. Here, $\Phi$ is the measurement matrix constructed using LWCDA algorithm and analyzed for different $\Psi$ (DCT, DFT, DWT, Laplacian and DiWT) matrices. (a) - (d) show values of $\delta_{k}$, $\lambda_{max}$ and $\lambda_{min}$ for matrix $A$ at different compression rates $\Gamma$. CS matrix $A$ gives a better range of $k$ with DFT and DCT compared to DWT basis in grid deployment. In the random deployment case, CS matrix $A$ gives slightly better range of $k$ with Laplacian compared DiWT.}
        \label{delk}       
\end{figure*}

\subsubsection{Numerical verification of $\delta_{k}$, $\lambda_{min}$ and $\lambda_{max}$}
The DFT, DCT, DWT, Laplacian and DiWT matrices are separately considered as bases $\Psi_{N \times N}$ for the empirical evaluation of $\delta_{k}$. To verify with DWT, we evaluated the performance of the CS matrix $A$ across several popular wavelets such as Daubechies, Symlets, Coiflets and chose the Daubechies-4 wavelet for all our analysis as it gives a better range for $k$ compared to other wavelets. The compressed sensing matrix $A_{M \times N} = \Phi_{M \times N}\Psi_{N \times N}$ with $N=1024$, at different compression rates with $M= 103, 308, 717, 922$ are considered for evaluation. For a compressed sensing matrix $A\in \mathbb{R}^{M \times N}$, the compression rate $\Gamma\%$ can be written as, $\Gamma \%= \Big(1-\dfrac{M}{N}\Big)\times 100.$

The procedure followed for empirical evaluation of $\delta_{k}$, $\lambda_{min}$ and $\lambda_{max}$ is described below:
\begin{enumerate}
\item Generate the measurement matrix $\Phi$ and the basis $\Psi$ for fixed $N=1024$ and for each $M$.  
\item For a combination $(N,M)$, $k$ is varied across $[1:M]$. 
\item Consider a $k$-sparse vector $u\in \mathbb{R}^{N}$. The vector $u$ contains non-zero values at $k$ randomly chosen locations and the values themselves are chosen from a normal distribution.
\item Find the support set for $u$, i.e., $T$.
\item Repeat steps 2 and 3 for 10000 iterations for each combination $(N,M,k)$ and calculate $\delta_{k}$ from \eqref{rip1}.
\item Compute $\lambda_{min}(A_{T}^{*}A_{T})$, $\lambda_{max}(A_{T}^{*}A_{T})$, where $T$ is the support set corresponding to $\delta_{k}$ from step 5. 
\end{enumerate}

The calculated RIC $\delta_{k}$ values, $\lambda_{max}$, $\lambda_{min}$ with respect to sparsity value $k$ at different compression rates $\Gamma$, are plotted in Fig.~\ref{delk}. In Fig.~\ref{delk}, $\delta_{kf}$, $\delta_{kc}$, $\delta_{kw}$, $\delta_{kl}$, $\delta_{kd}$ refer to RICs of CS matrix $A$ where $\Psi$ is DFT, DCT, DWT, Laplacian and DiWT respectively. $\lambda_{min}$, $\lambda_{max}$ refer to the minimum and the maximum eigenvalues of CS matrix $A$ respectively when $\Psi$ is DFT. Similar behavior of eigenvalues is also observed with DCT, DWT, Laplacian and DiWT bases.   

\begin{table}
\begin{center}
\caption{Sparsity value $k$ where RIC $\delta_{k}$ $\in (0,1)$ $\forall$ $u\in \Sigma_{k}$ for different $\Gamma$.}
\begin{tabular}{ | c  | c |  c | c | c | c |}
\hline
\multicolumn{6}{|c|}{Sparsity value $k$}\\
\hline
 \multicolumn{1}{|c|}{}&\multicolumn{3}{|c|}{Regular deployment} & \multicolumn{2}{|c|}{Random deployment} \\
 \hline
 Compression rate $\Gamma$ & DFT & DCT & DWT & Laplacian & DiWT\\
 \hline
 \hline 
 90\%   &   7   & 6 & 1  & 2 & 1 \\
 \hline 
 70\%   &   15   & 14 & 3 & 4 & 3\\
 \hline
 30\%   &   36  &   35 & 9 & 13 & 9\\
 \hline
 10\%   &   67  &   66 &  11 & 19 & 13\\
 \hline
\end{tabular}
\label{table}
\end{center}
\end{table}

Sparsity values $k$ obtained while $\delta_{k}\in (0,1)$ for the proposed CS matrix $A$ with different bases are tabulated in Table I (the same can be observed from Fig.~\ref{delk} as well).  
The interesting observation made from Table I is that the CS matrix $A$ gives better range for $k$ with DFT compared to that of DCT and DWT bases. In the random deployment case, CS matrix $A$ gives slightly better range for $k$ with Laplacian then DiWT basis.

\subsection{Coherence Analysis}
As discussed in Section \ref{cs}, if $X$ can be sparsely represented in an arbitrary basis $\Psi$, then for successful recovery, CS theory requires low mutual coherence between columns of the matrix $A=\Phi\Psi$.   
The mutual coherence $\mu$ of the matrix $A$ with different bases at various compression rates $\Gamma$ is calculated using \eqref{cohr}, i.e., the CS matrix $A_{M \times N} = \Phi_{M \times N}\Psi_{N \times N}$ where $N=1000$ and $M$ is chosen to vary from $100$ to $900$ in steps of $100$ ($M= 100:100:900$) for calculating $\mu$.
      
\begin{figure}[ht]
\centering
\includegraphics[scale=0.4]{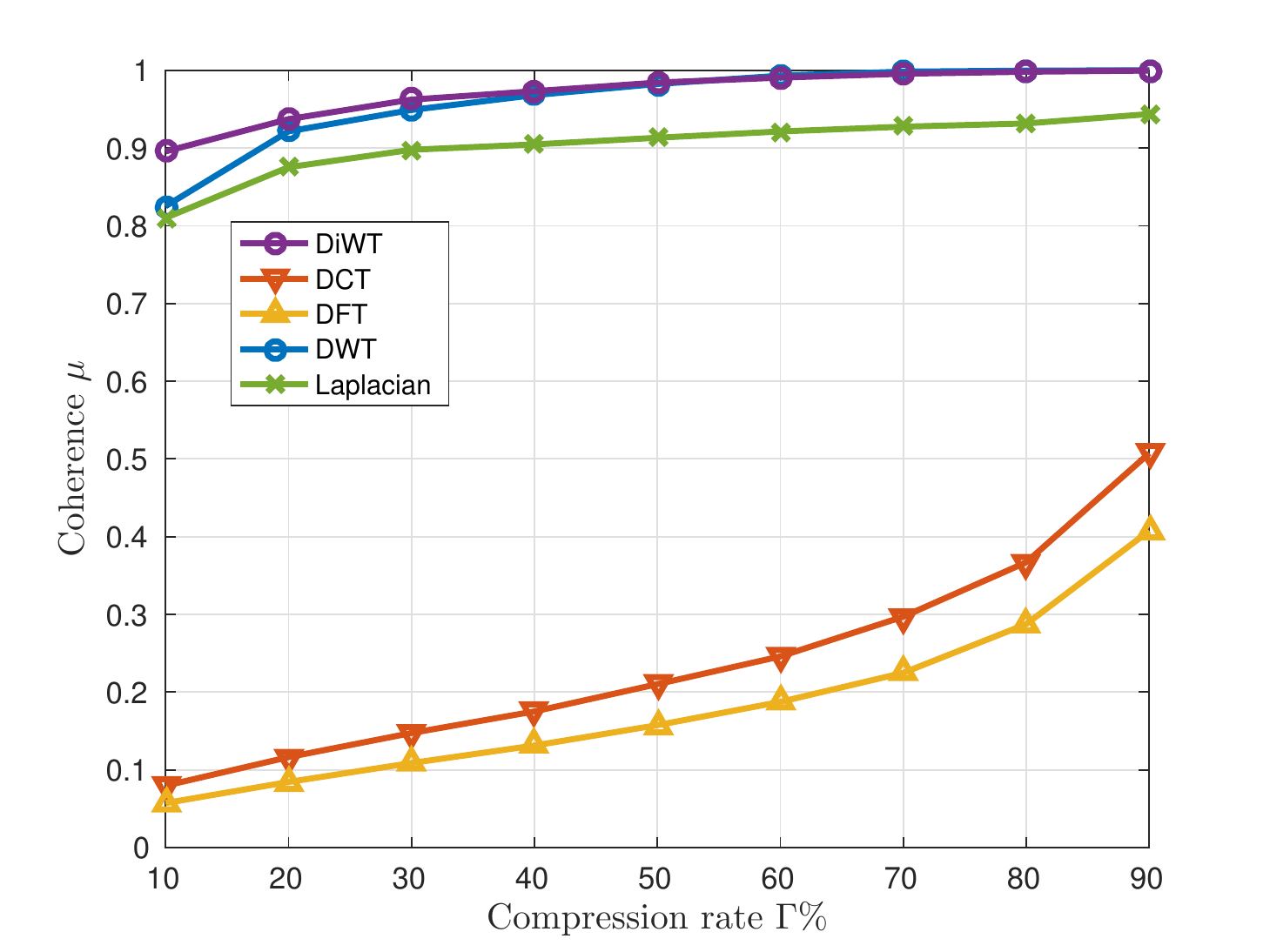}
        \caption{Comparison of mutual coherence $\mu$ of the CS matrix $A=\Phi\Psi$ with different bases where matrix $\Phi$ is constructed from LWCDA. Columns of the matrix $A$ are highly incoherent with DFT basis among all.}
        \label{cohr_plt}       
\end{figure}

The resultant mutual coherence with different bases are shown in Fig.~\ref{cohr_plt}. The CS matrix $A$ provides better incoherence for the DCT and DFT bases compared to the DWT basis where $\Phi$ is constructed from grid deployment. In case of random deployment, the coherence of the matrix $A$ with Laplacian is fairly better compared to DiWT basis across all compression rates. It is observed from Fig.~\ref{cohr_plt} that among all the bases, DFT provides high incoherence for all compression rates. 

\subsection{Phase Transition Analysis}
\label{pt}

\begin{figure*}[ht]
	\begin{subfigure}[b]{0.2\textwidth}
	\centering		
		\includegraphics[width=\linewidth]{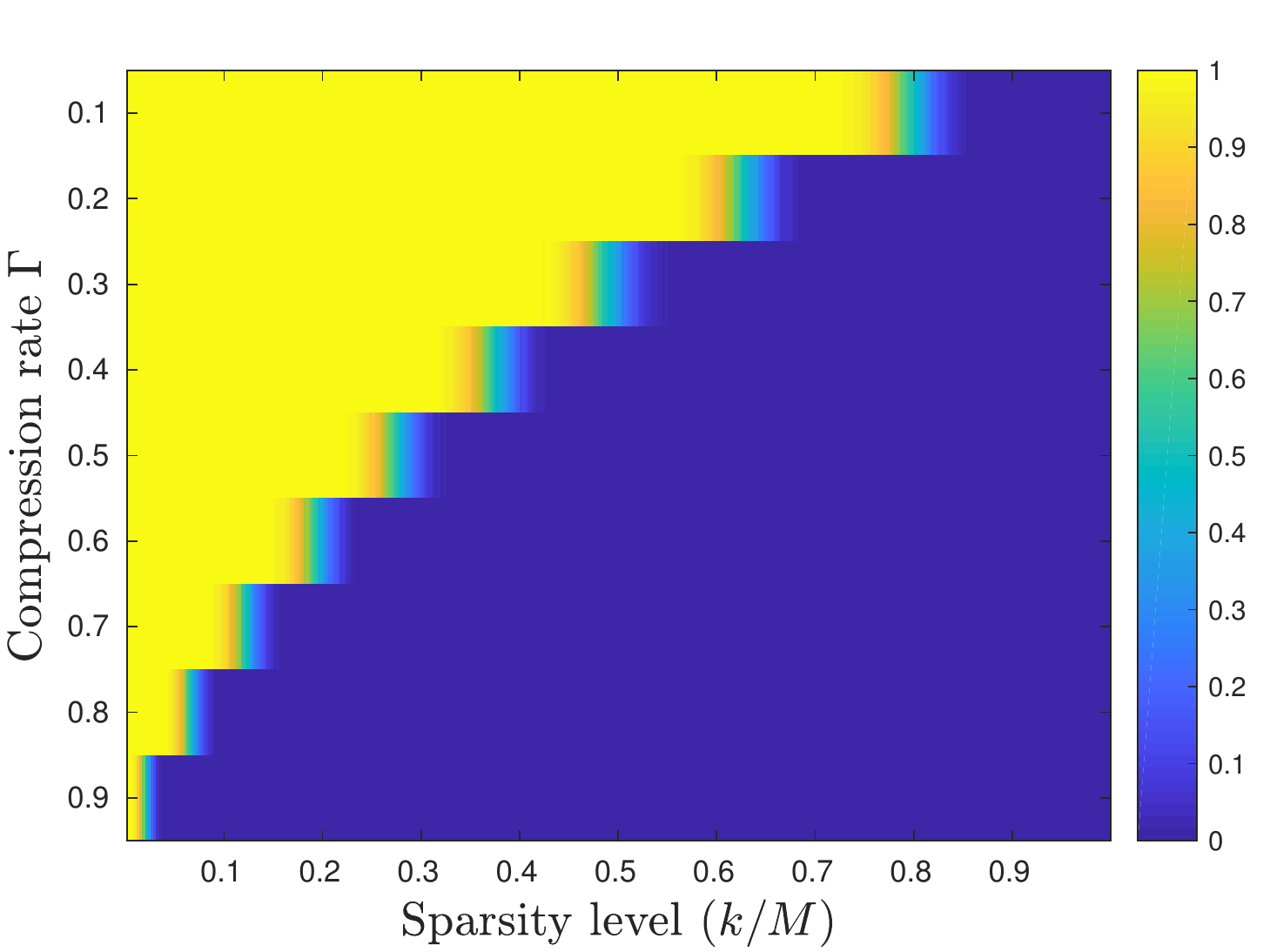}
		\caption{DCT.}
		\label{Pt_grid_dct}
	\end{subfigure}%
	\begin{subfigure}[b]{0.2\textwidth}
	\centering	
		\includegraphics[width=\linewidth]{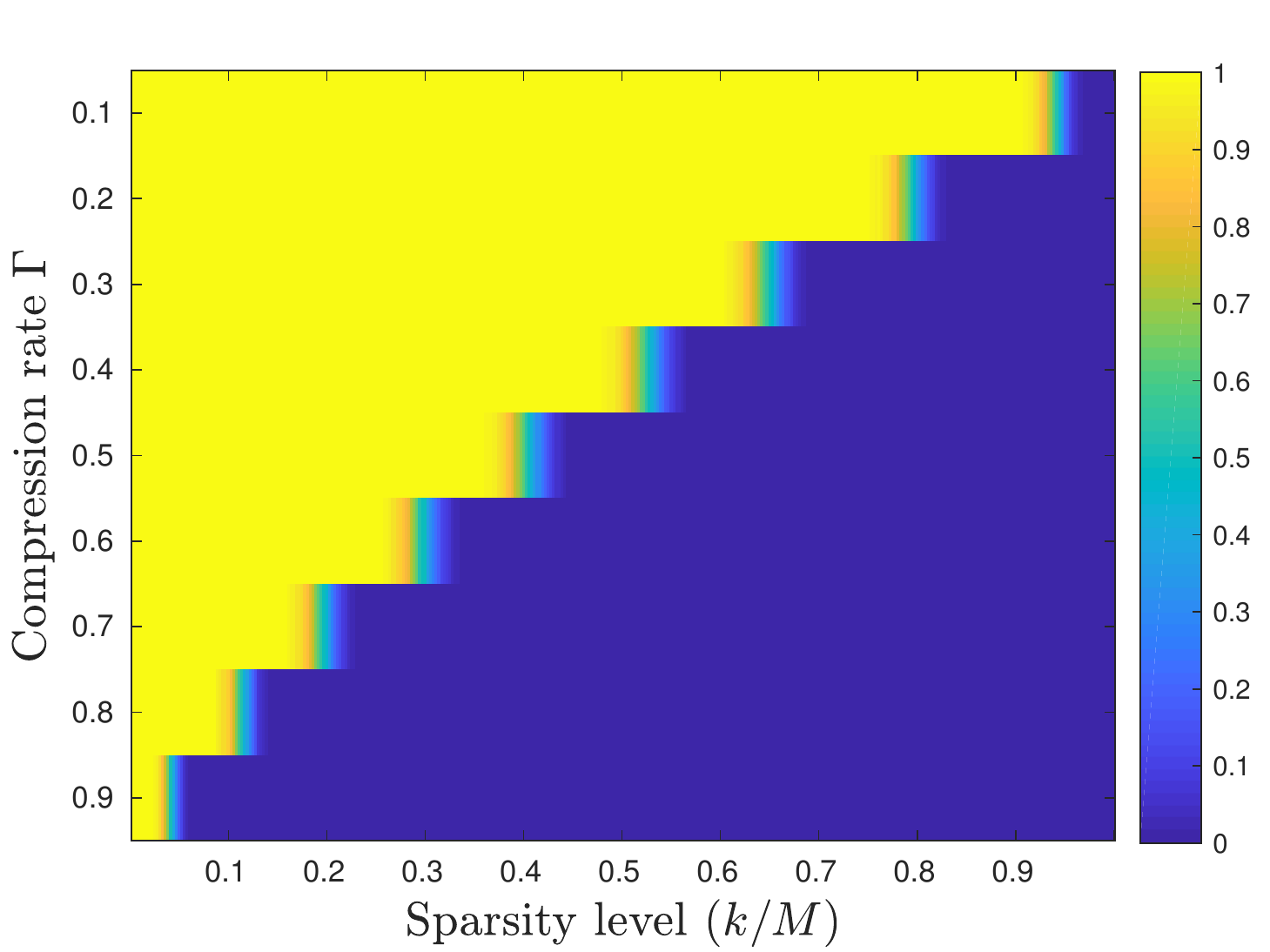}
		\caption{DFT.}
		\label{Pt_grid_dft}
	\end{subfigure}%
	\begin{subfigure}[b]{0.2\textwidth}
	\centering
		\includegraphics[width=\linewidth]{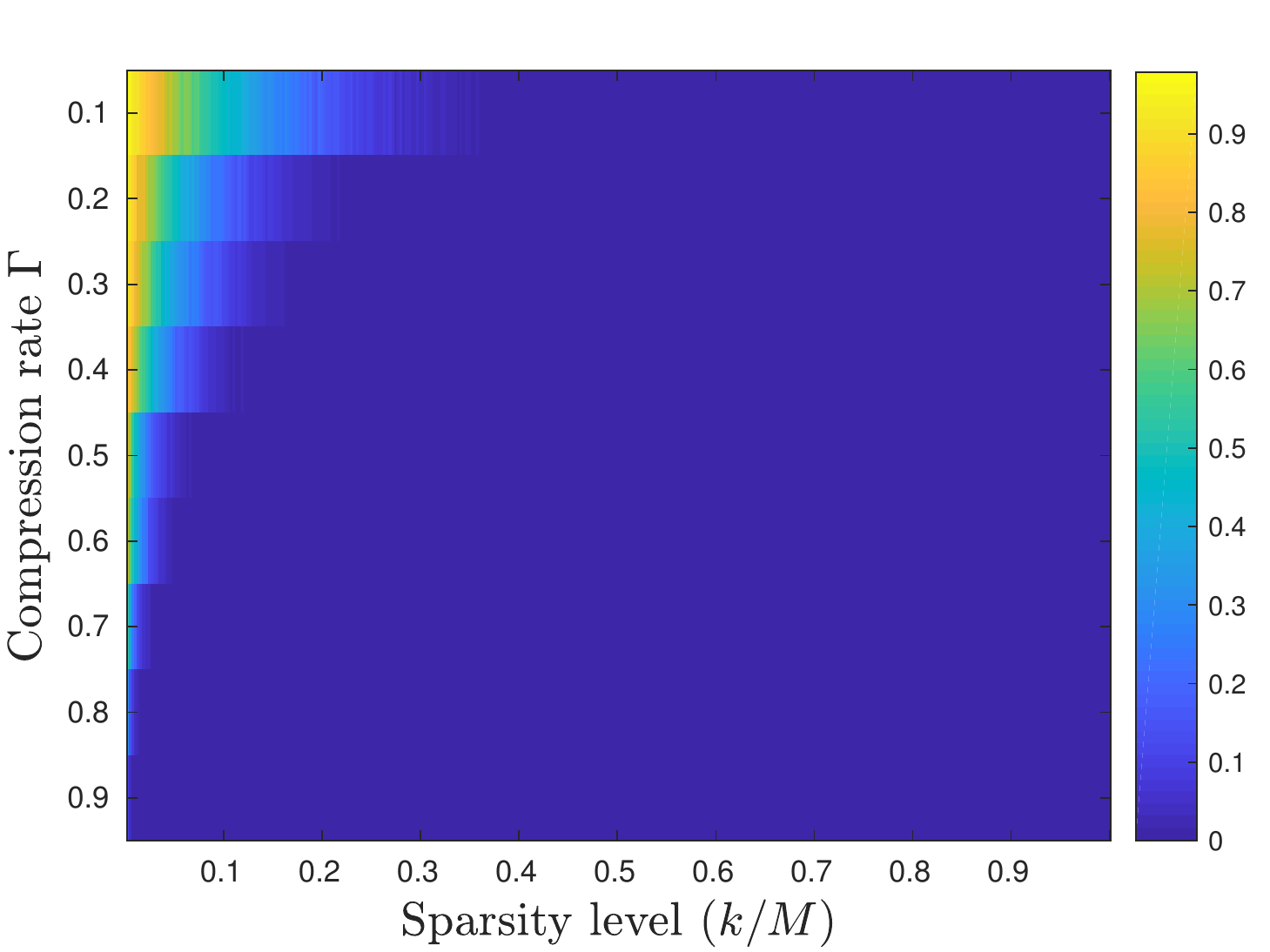}
		\caption{DWT.}
		\label{Pt_grid_dwt}
	\end{subfigure}%
	\begin{subfigure}[b]{0.2\textwidth}	
	\centering
		\includegraphics[width=\linewidth]{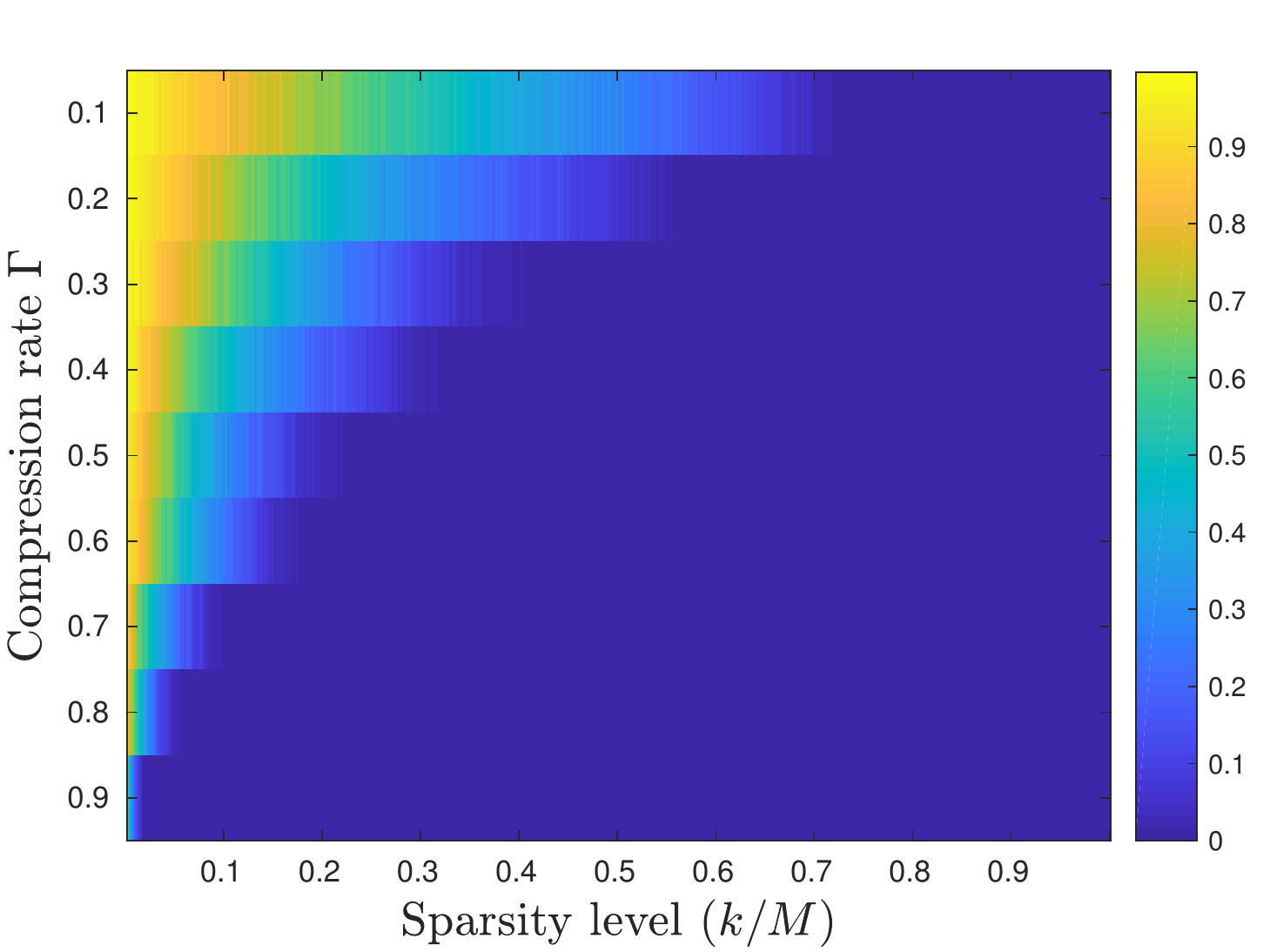}
		\caption{Laplacian.}
		\label{Pt_grid_dwt}
	\end{subfigure}%
	\begin{subfigure}[b]{0.2\textwidth}	
		\includegraphics[width=\linewidth]{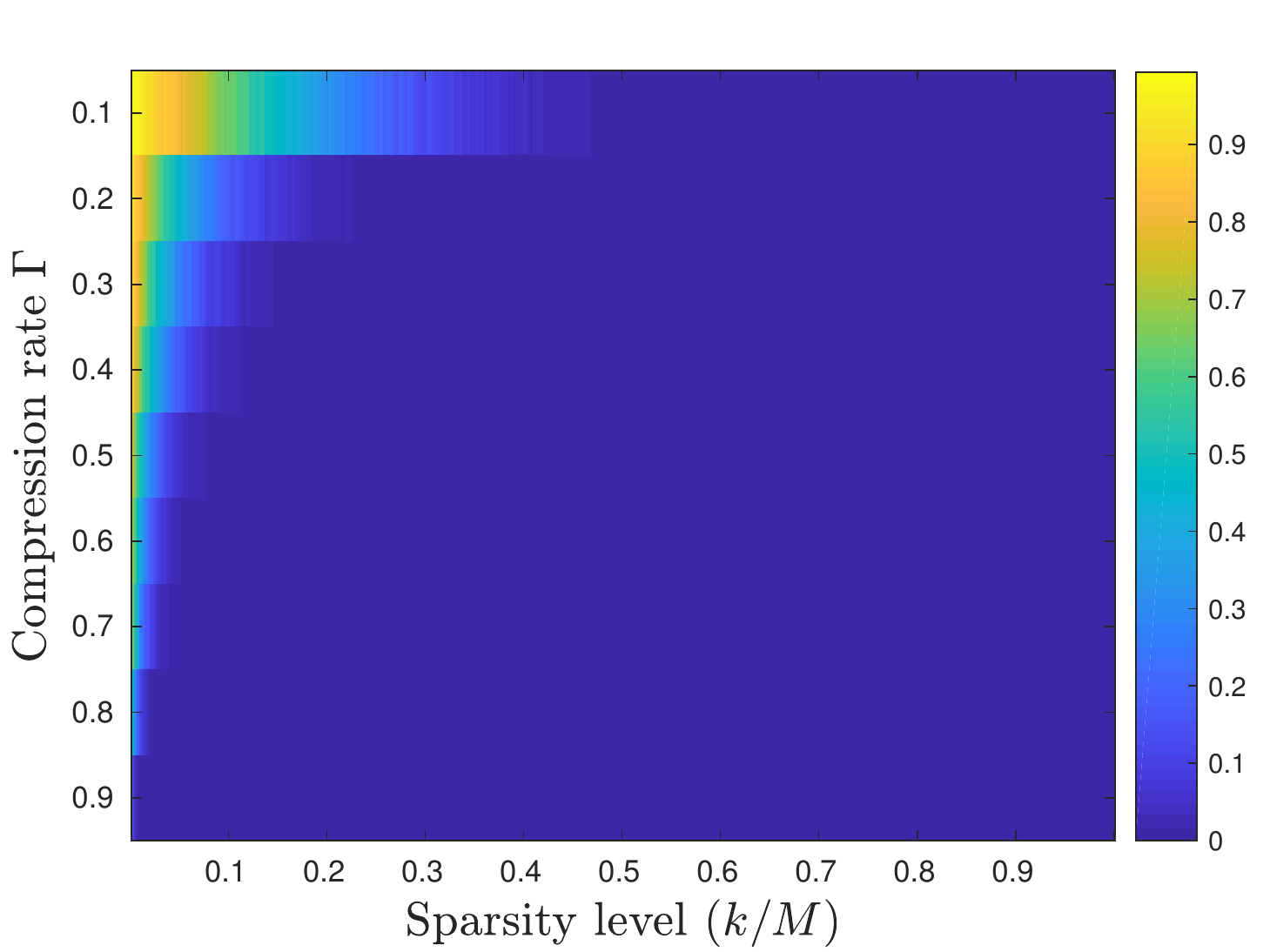}
		\caption{DiWT.}
		\label{Pt_grid_dwt}
	\end{subfigure}%
	\caption{Phase transition analysis of CS matrix $A=\Phi\Psi$ for different bases, where $\Phi$ is the proposed measurement matrix and $\Psi$ is the basis matrix. The color bar indicates successful recovery probability $P_{s}$. CS matrix $A$ with DCT and DFT basis yields promising recovery region.}
	\label{pt_analysis}       
\end{figure*}
For a given CS matrix, the phase diagram can be generated as a numerical representation of successful recovery probability $P_{s}$ over the space $(k/M,1-(M/N)) \in [0, 1]^{2}$, as in \cite{Pt}. This space is discretized and we performed multiple compression and decompression experiments at each grid point. The phase diagram is finally approximated by using successful recovery probability $P_{s}=Pr\{e \leq e_{TH}\}$, where the reconstruction error $e=\norm{X-\hat{X}}/\norm{X}$, with an appropriately selected threshold $e_{TH}$. We considered error threshold $e_{TH}=10^{-8}$ in our analysis. For PT analysis, $A_{M \times N} = \Phi_{M \times N}\Psi_{N \times N}$ is considered with $N=1000$ and evaluated for different compression rates ($\Gamma$) with $M= 100:100:900$. Fig.~\ref{pt_analysis} shows the phase diagram of CS matrix $A=\Phi\Psi$, where $\Phi$ is the measurement matrix and $\Psi$ is the basis. Fig.~\ref{pt_analysis} also illustrates that the proposed measurement matrix $\Phi$ with DCT and DFT bases provides promising recovery region compared to DWT where $\Phi$ is constructed from grid deployment. In case of random deployment, Laplacian basis provides slightly better recovery region compared to DiWT basis.

This evaluation has shown that the proposed measurement matrix $\Phi$ gives better performance with DCT and DFT bases compared to the DWT basis in terms of RIC, coherence and PT analysis where $\Phi$ is designed from grid-wise deployed network. Further, in random deployment scenario, Laplacian and DiWT bases give comparable performance. The proposed matrix $\Phi$ with DCT and DFT bases ($\Psi$) has the ability to recover the signals successfully even though they have fairly low sparsity. Whereas in case of DWT, Laplacian and DiWT, the matrix $\Phi$ can recover the signals on the condition that they are highly sparse.

To extend the proposed LWCDA method to fairly low sparse data cases especially in random deployment scenario, we propose a technique called spatial logical node mapping, which is described in the following subsection.

\subsection{Spatial Logical Node Mapping}

\begin{figure}[ht]
\centering
	\begin{subfigure}[b]{0.25\textwidth}		
		\includegraphics[width=\linewidth]{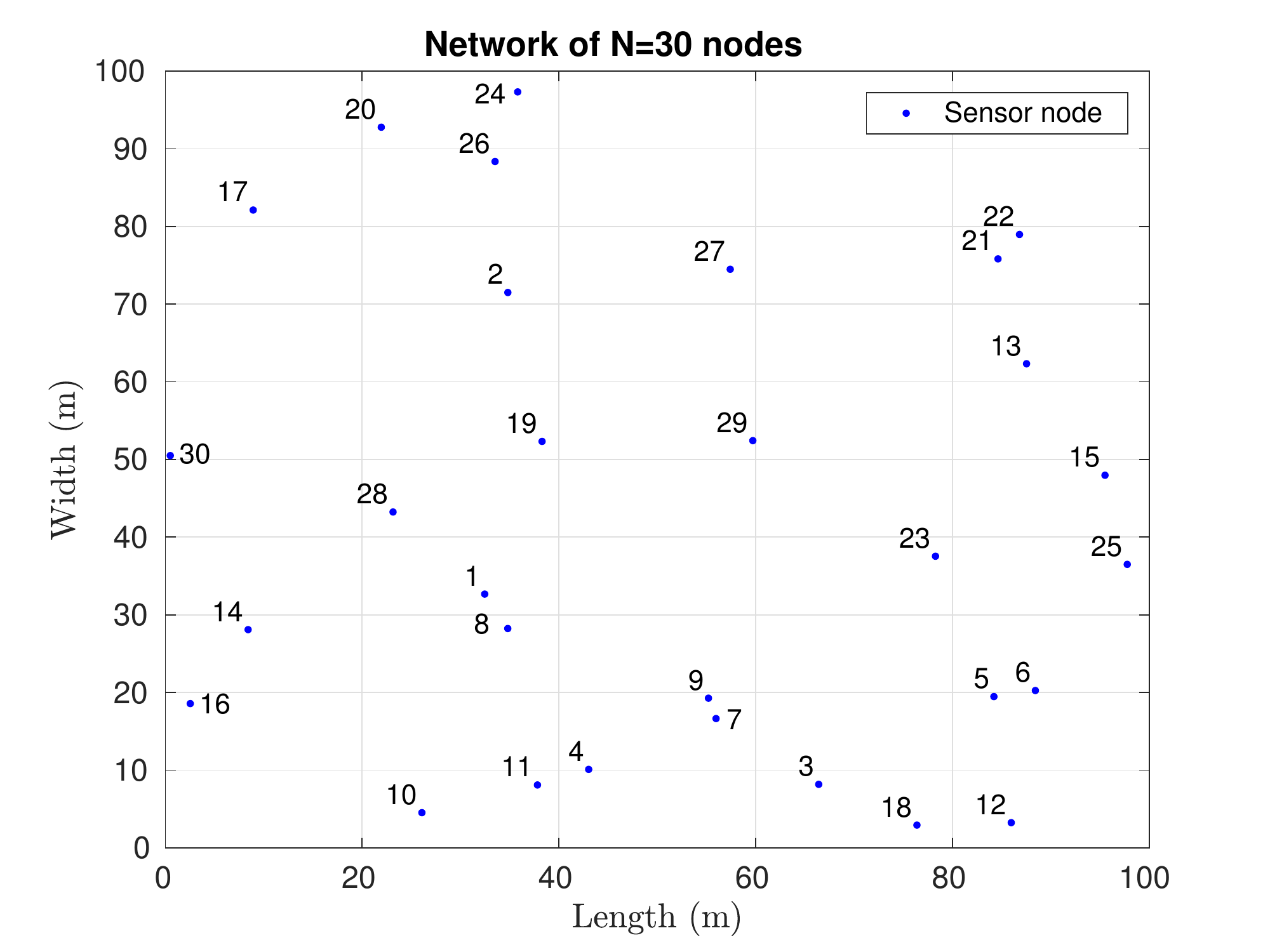}
		\caption{Randomly deployed nodes.}
		\label{ndmpnga}
	\end{subfigure}%
\begin{subfigure}[b]{0.25\textwidth}
		\includegraphics[width=\linewidth]{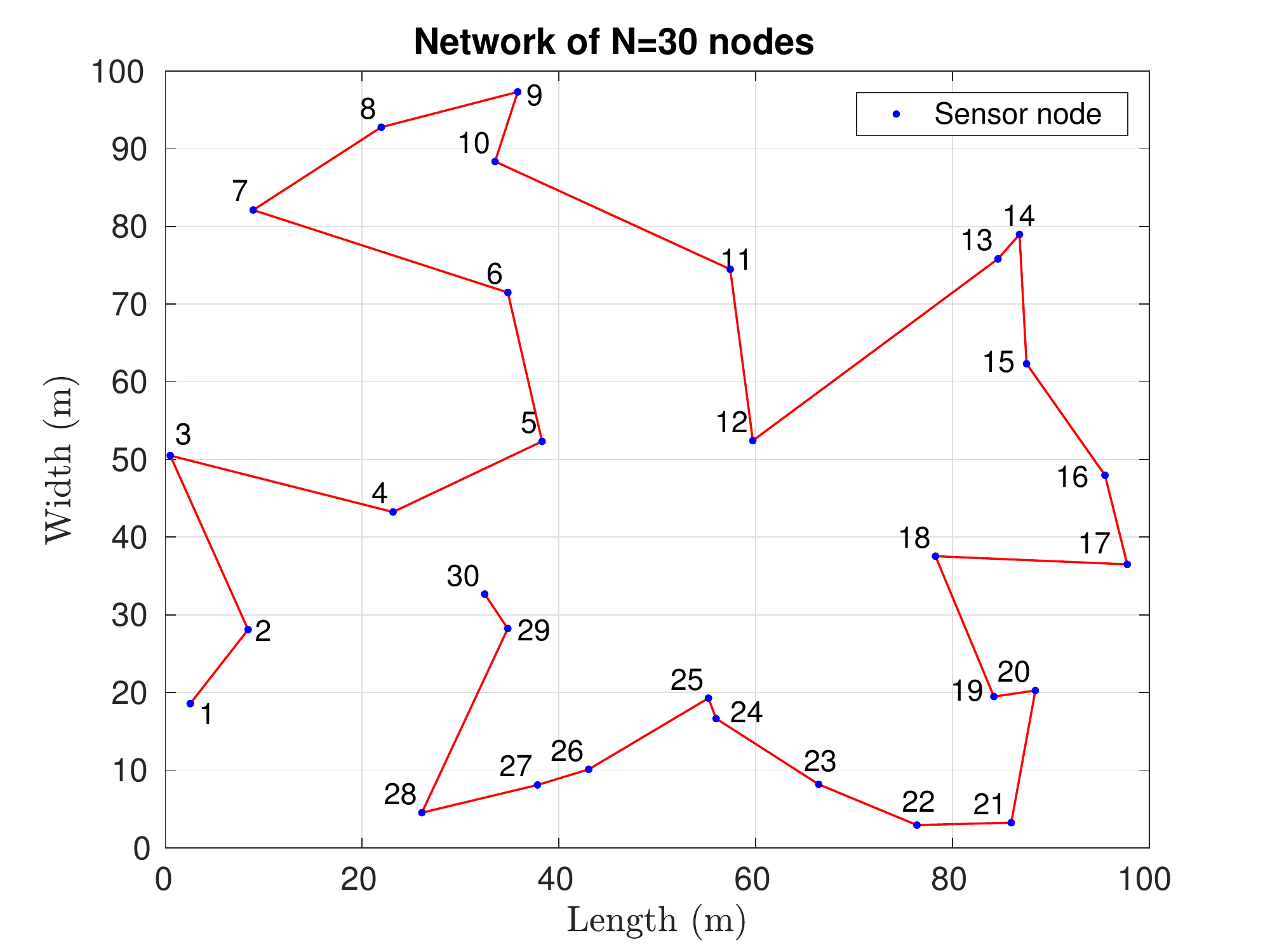}
		\caption{Logical mapped nodes.}
		\label{ndmpngb}
	\end{subfigure}
	\caption{Logical node mapping.}
	\label{ndmpng}       
\end{figure}

Before invoking the LWCDA algorithm, we first model the network as a logical chain based on the Euclidean distance between the nodes. The algorithm starts from any random node and gives sequential node IDs along the chain. The method used to form the logical chain is similar to that in \cite{lg1}. We consider that in the initial phase, each node sends the distance information of the nodes that are in its coverage  area to the sink. The sink maps the new node IDs from old node IDs and sends it back to the nodes to change. Fig. \ref{ndmpng} provides more insight into the Spatial Logical Node Mapping (SLNM) with an example network of $N=30$ nodes. This preprocessing will introduce spatial correlation in the data since adjacent nodes in the chain tend to be the nodes which are geographically close to each other \cite{lg1} \cite{lg3}. The spatial correlation among the samples generated from the nodes which are geographically close to each other can make the signal sparse in the regular DFT and DCT bases. SLNM adds the advantage to LWCDA to guarantee the recovery of the measured data from the random deployment as it introduces sparsity for the data in DFT and DCT bases. 

In the following section, we evaluate the performance of the proposed data aggregation method in an application scenario.  

\section{Results And Evaluation}
In this section, the performance of the proposed data aggregation method is analyzed using the following metrics:
\begin{enumerate}
\item[1.] Reconstruction error.
\item[2.] Transmission cost.
\end{enumerate}
\subsection{Reconstruction Error Analysis}
\begin{figure}[!t]
\centering
\includegraphics[scale=0.4]{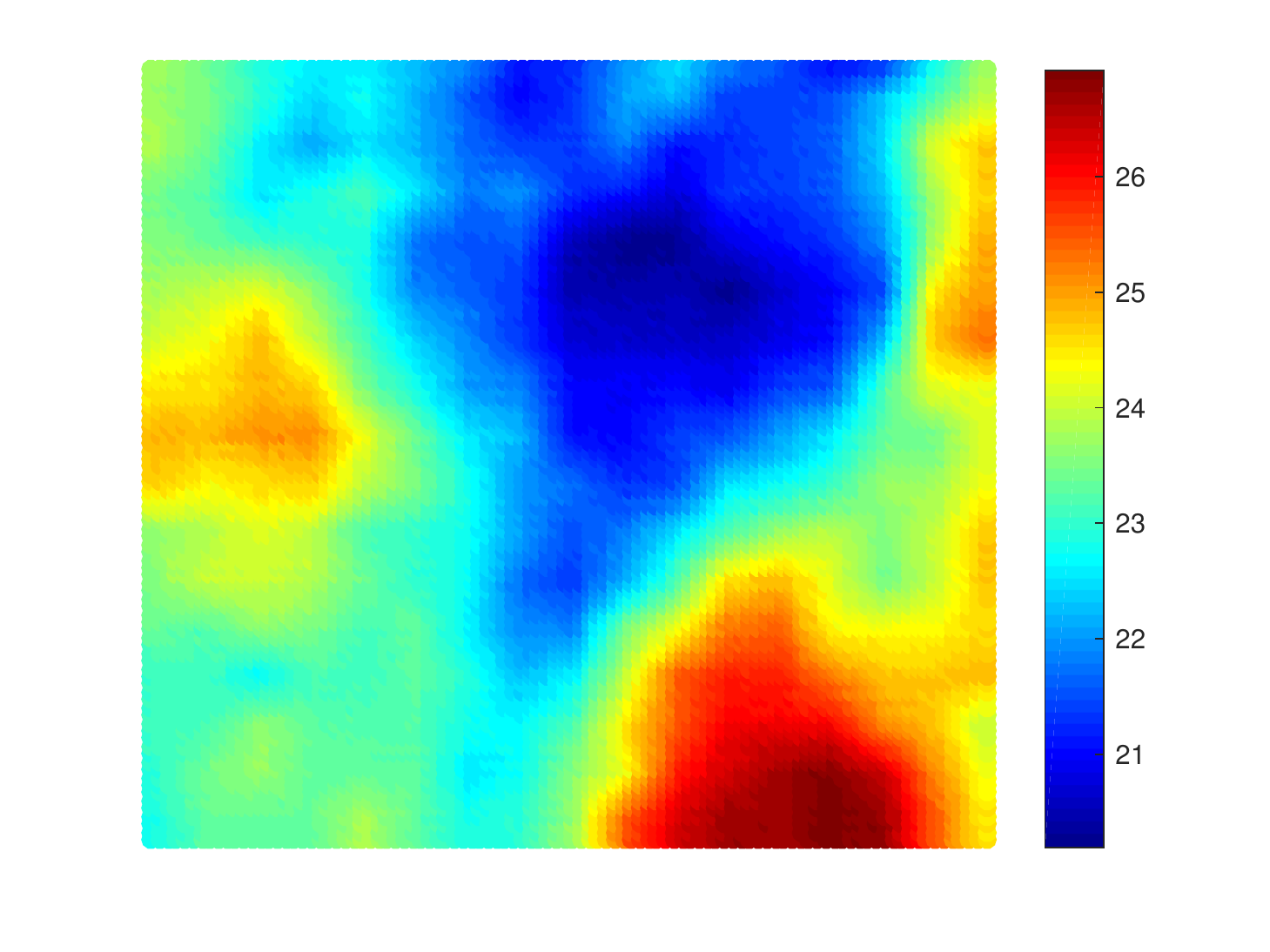}
\caption{Temperature sensing field.}
\label{temp}       
\end{figure}

\begin{figure}[!t]
\centering
\includegraphics[scale=0.4]{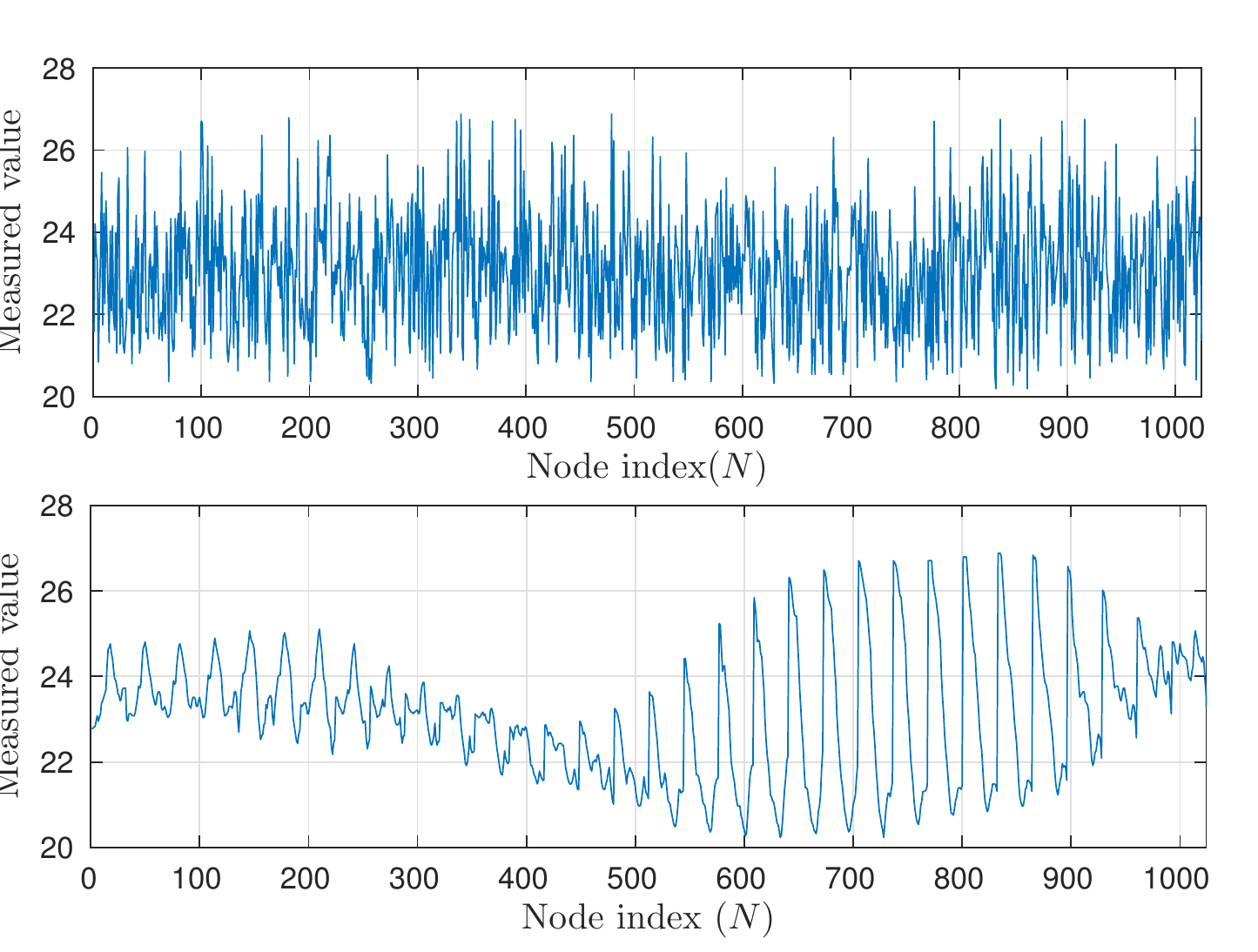}
        \caption{Measured data from random (top) and grid-wise (bottom) deployed 1024-node network.}
        \label{sensing_data}       
\end{figure}
We extend the application of the proposed algorithm for compressible signals by using a real data set for evaluation. Real temperature data which is obtained by capturing thermal images from the top view of an area $100 m \times 100 m$ is considered for analysis. Fig.~\ref{temp} visualizes the considered temperature data for recovery performance of the proposed method. Fig.~\ref{sensing_data} shows the measured data from random and grid-wise deployed 1024 sensor nodes on the field. We used MATLAB R2015b software for performing all our simulations. Ideally the sparsity value $k$ of $X$ in a basis $\Psi$ is measured using the $l_{0}$ norm, $k=\|{\theta}\|_{0}$, where $\theta=\Psi X$. For real-time data which is approximately sparse, only few large coefficients contribute a large proportion of the total energy. We use numerical sparsity \cite{diwt_basis} as the measure of sparsity which represents the number of effective large coefficients. If a vector $X$ can be represented using a sparsifying basis $\Psi$ as $X=\Psi \theta$, then the numerical sparsity of $X$ can be calculated as
\begin{equation}
s=\frac{{\|\theta\|_{1}}^{2}}{{\|\theta\|_{2}}^{2}}.
\end{equation}
Numerical sparsity of the considered temperature data with different bases ($\Psi$) (DFT, DCT, DiWT and Laplacian) are tabulated in Table II. 
 
\begin{figure}[ht]
\centering		
	\begin{subfigure}[b]{0.25\textwidth}
		\includegraphics[width=\linewidth]{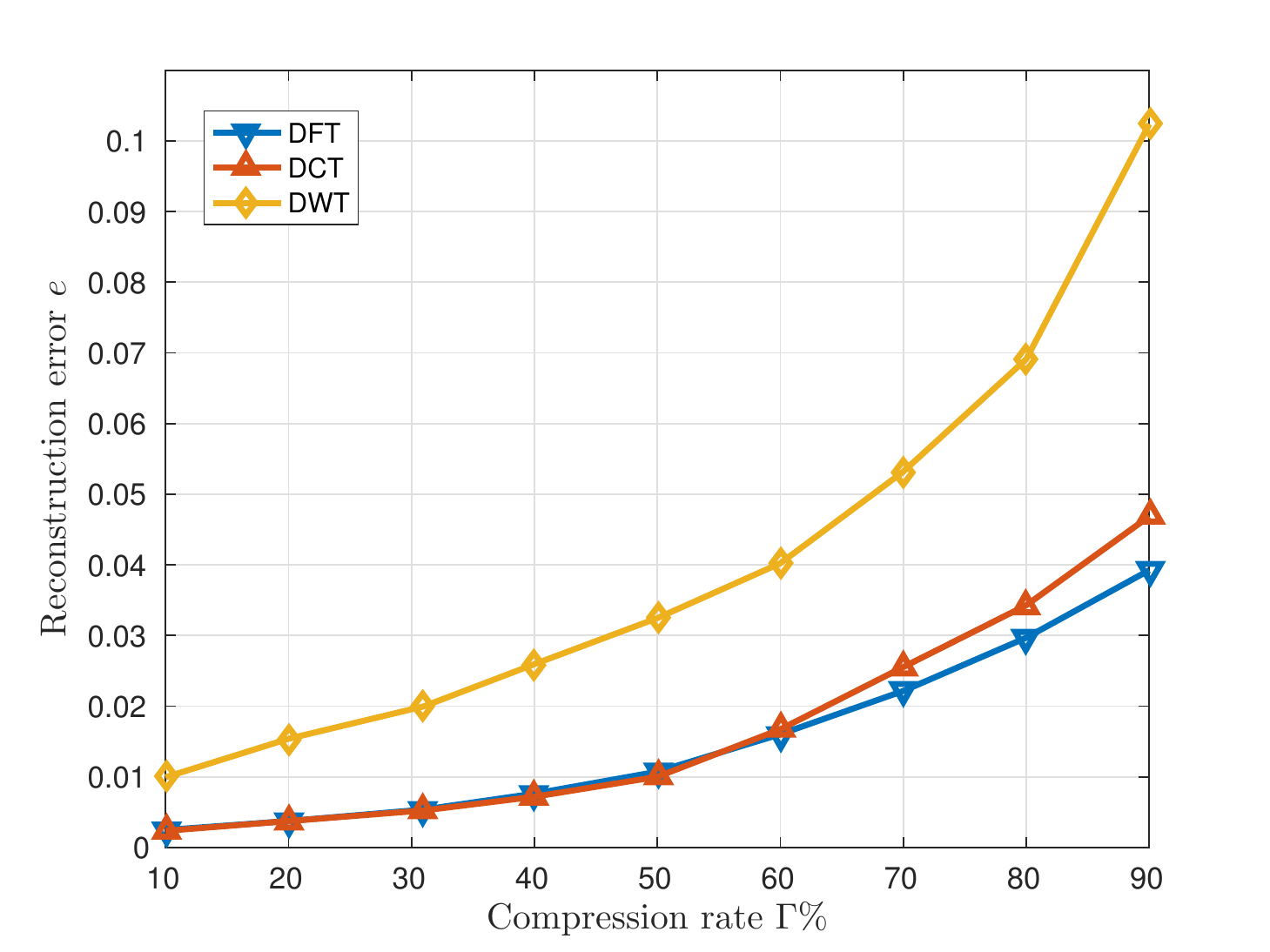}
		\caption{}
		\label{grid_recovery}
	\end{subfigure}%
	\begin{subfigure}[b]{0.25\textwidth}
		\includegraphics[width=\linewidth]{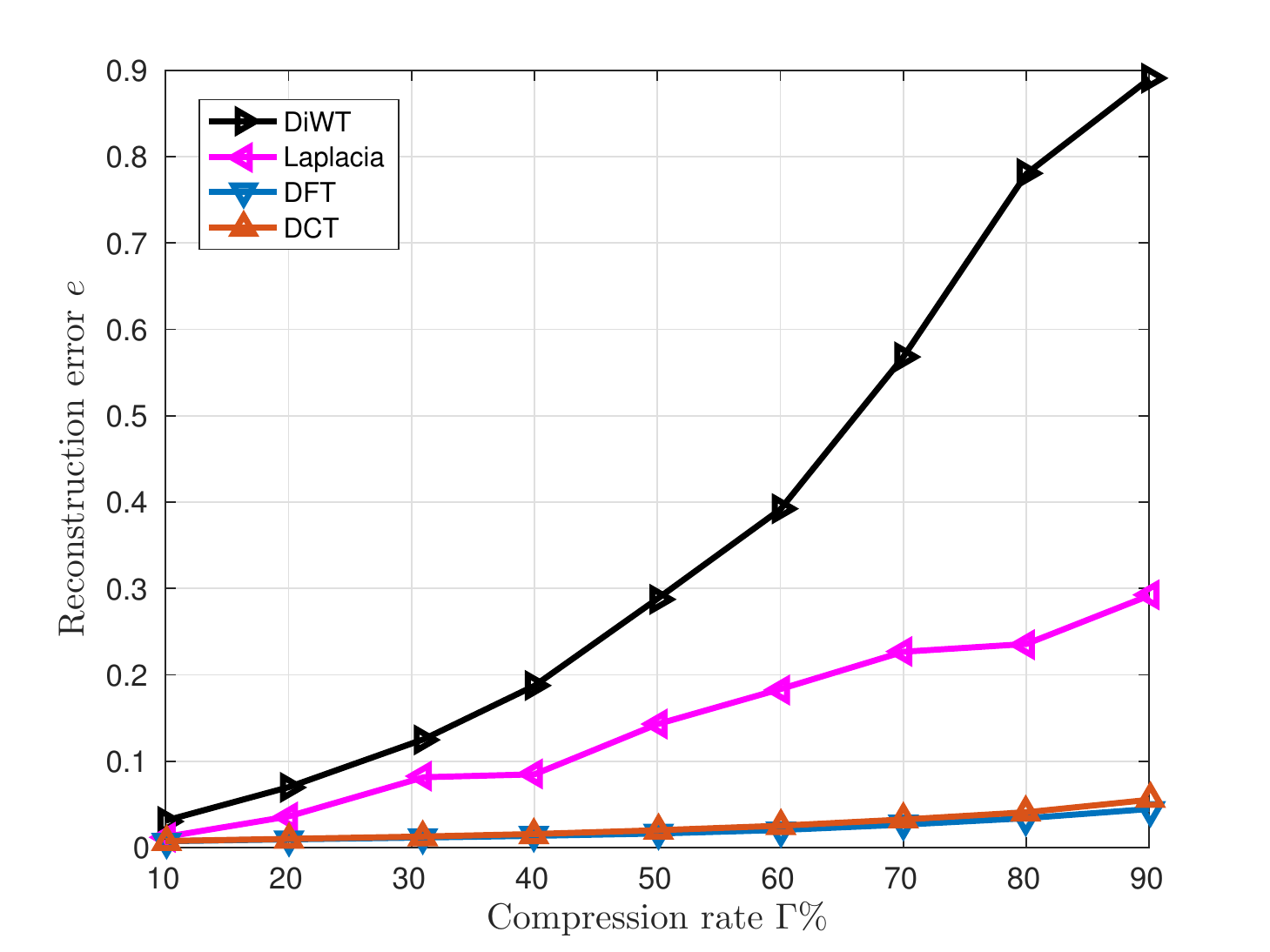}
		\caption{}
		\label{random_recovery}
	\end{subfigure}
	\caption{Average reconstruction error against different compression rates. Here, 1024 temperature data points are considered. Figures (a) and (b) depicts recovery of grid-wise and Randomly deployed nodes measured data respectively. DFT basis provides low recovery error compared to all bases in both grid and random deployment scenario.}
	\label{error}       
\end{figure}
 
Table II shows considered data is more sparse in DCT and DFT bases compared to the others in both grid-wise and random deployment scenarios. To evaluate recovery performance of the measured data from grid-wise deployed nodes the basic LWCDA is used to construct the measurement matrix $\Phi$. In case of random deployment, $\Phi$ is constructed from LWCDA and SLNM. The OMP algorithm \cite{OMP} is used for the recovery of the compressed data. We evaluated the performance of the proposed data aggregation method in terms of the reconstruction error $e$ against the compression rate $\Gamma$. Fig.~\ref{error} compares average reconstruction error of our method with different bases. In our analysis, $e$ is averaged over 100 iterations for each $\Gamma$. From Fig.~\ref{grid_recovery}, it can be observed that DFT and DCT can recover the data which is measured from grid-wise deployed nodes with a low error compared to DWT for all compression rates. Fig.~\ref{random_recovery} illustrates the data recovery performance of the proposed method where the data is measured from randomly deployed sensor nodes. From Fig.~\ref{random_recovery}, it can be observed that DCT and DFT can recover the data with a low error across all compression rates compared to other bases. However, DiWT and Laplacian result in high recovery error as they require the signal to be highly sparse. This evaluation has shown that the proposed LWCDA method provides high recovery fidelity using the DFT basis for the data measured from both the random and grid-wise deployed nodes.

\begin{table}
\begin{center}
\caption{Sparsity measure of the temperature data}
\begin{tabular}{ | c | c |  c | c | c | c | c |}
\hline
\multicolumn{7}{|c|}{Numerical sparsity value $s$}\\
\hline
 \multicolumn{3}{|c|}{Regular deployment}&\multicolumn{4}{|c|}{Random deployment}\\
 \hline
 \multicolumn{3}{|c|}{}&\multicolumn{2}{|c|}{}&\multicolumn{2}{|c|}{SLNM}\\
 \hline
 \hline
 DFT & DCT & DWT & Laplacian & DiWT & DFT & DCT \\
 \hline
 2.2205  &  2.5251  & 7.6707 & 15.925  & 53.0402 & 2.6219 & 2.7569 \\
 \hline
\end{tabular}
\label{table}
\end{center}
\end{table}

In the following section, we perform a comparative analysis of the transmission cost of our algorithm with traditional CS based data gathering methods. To demonstrate the efficiency of our algorithm, we compare with SPRM for the grid-wise deployment scenario, CWCDA, Hybrid CS and Non-CS methods for the random deployment scenario. 

\subsection{Transmission Cost Analysis}
Transmission cost of the network $G(V,E)$ is defined as \cite{cso},
\begin{equation}
T_{cst}=\sum_{(i,j) \in E} {{t_{ij}c_{ij}}},
\end{equation}
where $t_{ij}$ represents the traffic on the link $(i,j) \in E$ and $c_{ij}$ is the cost of the link. We considered one packet as one unit of traffic on the link and cost of the link $c_{ij}$ is considered as the Euclidean distance between the nodes $i$ and $j$. ZigBee protocol is considered for simulations as the ZigBee stack is one of the most commonly used protocols among commercially available off-the-shelf IoT solutions. The size of PHY layer data field of the packet of ZigBee is 128 bytes, of which 87 bytes can be used for application payload as the remaining octets are reserved for packet header information of higher layers. The number of bits required to represent the data sample and the address field (short address mode) is considered to be 2 octets.

\begin{figure}
\centering
\includegraphics[scale = 0.48]{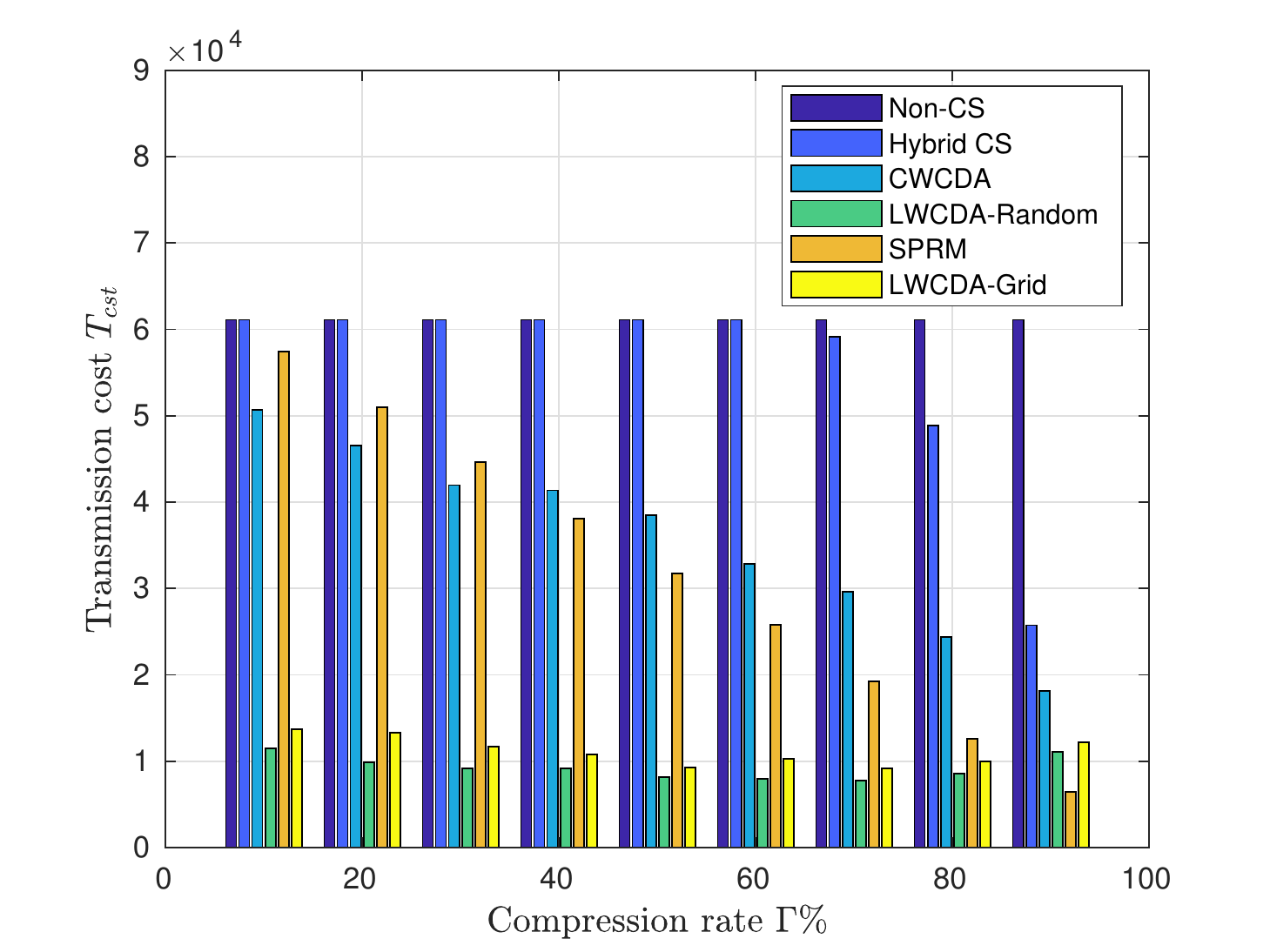}
\caption{Comparison of the transmission cost required for data aggregation from $625$ nodes deployed in an area $256m \times 256m$ using the proposed LWCDA and SPRM, Hybrid CS, CWCDA and Non-CS methods against the compression rate $\Gamma$. Transmission cost of the LWCDA is significantly low compared to all methods almost for all compression rates ($\Gamma \leq 80\%$).}
\label{tcost}
\end{figure}

\begin{figure}
\centering
\includegraphics[scale = 0.48]{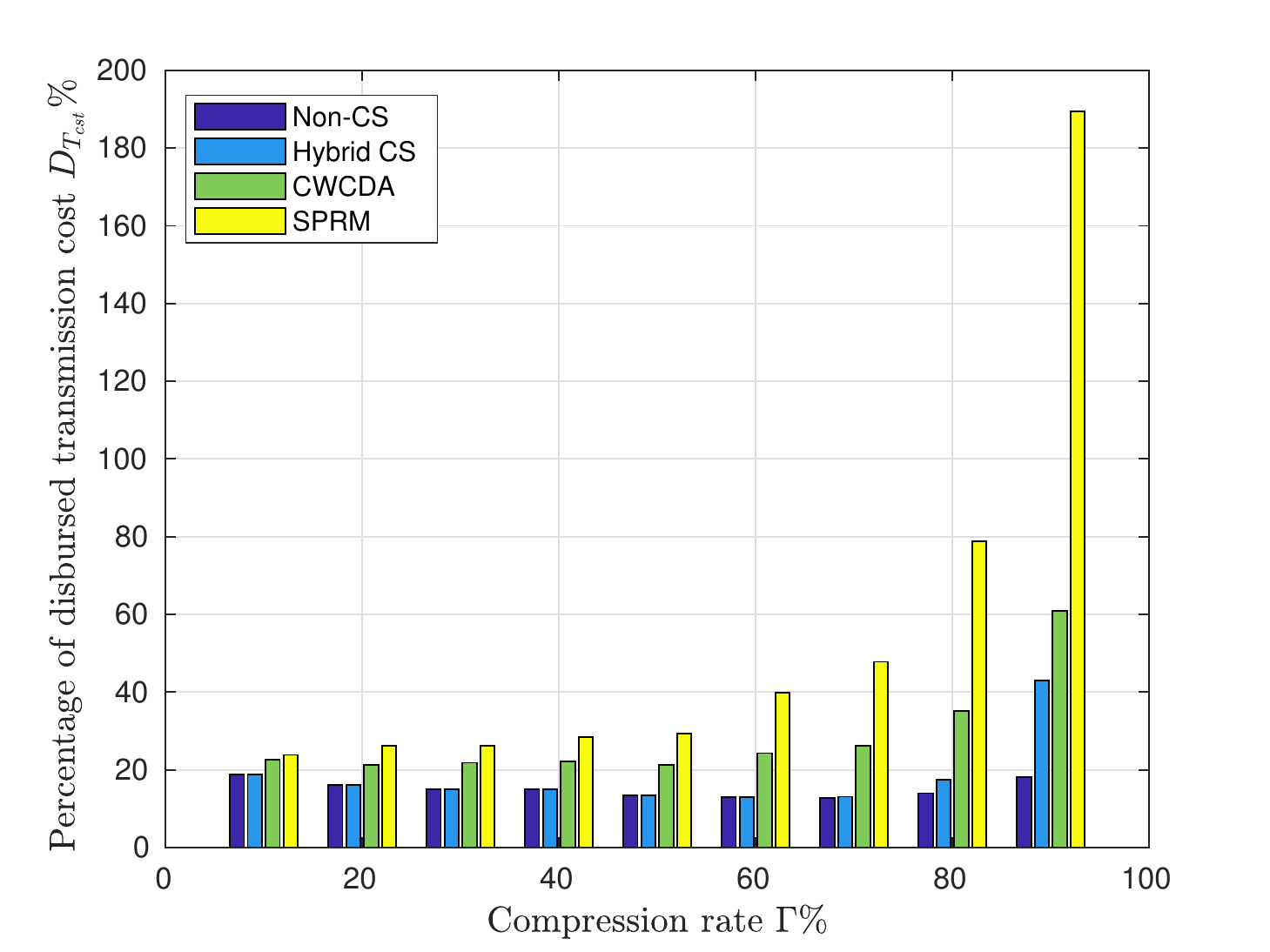}
\caption{Comparison of the percentage of disbursed transmission cost of the proposed LWCDA with respect to SPRM, Hybrid CS, CWCDA and Non-CS methods against the compression rate $\Gamma$. Here, data aggregation is considered from $625$ nodes deployed in an area $256m \times 256m$. Percentage of disbursed transmission cost of the LWCDA with respect to SPRM, Hybrid CS, CWCDA and Non-CS methods is low for almost all compression rates ($\Gamma \leq 80\%$).}
\label{tcost_save}
\end{figure}

For transmission cost comparison, a network deployment of $625$ nodes in an area of $256m \times 256m$ is considered. The comparison of the transmission cost for data aggregation using the proposed method (LWCDA), SPRM, Hybrid CS, CWCDA and a Non-CS method with respect to the change in compression rate ($\Gamma$) is shown in Fig.~\ref{tcost}. In particular, we considered the conventional shortest path algorithm \cite{geograph} for data gathering as the Non-CS approach, where each node in the network sends its data to the sink through the shortest path. From Fig.~\ref{tcost}, it can be observed that our LWCDA (labeled as LWCDA-Random for random deployment) method requires very low transmission cost for data aggregation compared to Non-CS, Hybrid CS and CWCDA for all compression rates where nodes are randomly deployed. In case of grid deployment, compared to SPRM the proposed LWCDA (labeled as LWCDA-Grid for grid deployment) method outperforms until $\Gamma=80\%$ for data aggregation. In the proposed LWCDA method, an increase in compression rate $\Gamma$ decreases the number of required clusters for data aggregation. A decrease in number of clusters increases the required transmission cost for data aggregation as the leaf nodes need to send their measurements to CHs from farther distance. Further, as $\Gamma$ increases, the required transmission cost  to collect measurements from CHs (using MST) also decreases. This results in an increase of total transmission cost $T_{cst}$ at higher compression rates ($\Gamma>80\%$) as shown in Fig.~\ref{tcost}.  Fig.~\ref{tcost_save} illustrates the percentage of disbursed transmission cost $D_{T_{cst}}$ of the proposed LWCDA with respect to that of Hybrid CS, LWCDA, SPRM and Non-CS methods. Percentage of disbursed transmission cost $D_{T_{cst}}$ of a given method $P$ with respect to the method $Q$ is defined as,
\begin{equation}
D_{T_{cst}}\%  =\frac{T_{cst}\text{ of method }P}{T_{cst}\text{ of method }Q}\times 100.
\end{equation}
The SPRM method at high compression rates ($\Gamma>80\%$) results in lesser transmission cost as compared to that of the proposed LWCDA method. This in turn results in the percentage of disbursed transmission cost of LWCDA (LWCDA-Grid) to go beyond $100\%$ as shown in Fig.~\ref{tcost_save} for higher compression rates. This is because, in the SPRM method, very few randomly selected nodes are required to send data through the shortest path to the sink at high compression rates. Although, SPRM offers higher compression rates with lower transmission costs, it does not achieve good performance with respect to coherence leading to higher reconstruction errors at higher compression rates, thereby not guaranteeing a successful reconstruction (as discussed in \cite{rout.5}). For all compression rates in both grid-wise ($\Gamma \leq 80\%$) and random deployment scenario, the proposed LWCDA method can deliver the data to the sink with a lower transmission cost as illustrated in Fig.~\ref{tcost} and with a lower percentage of disbursed transmission cost as shown in Fig.~\ref{tcost_save}, thereby enhancing the network lifetime as compared to the considered baseline approaches.

\begin{figure}
\centering
\includegraphics[scale = 0.48]{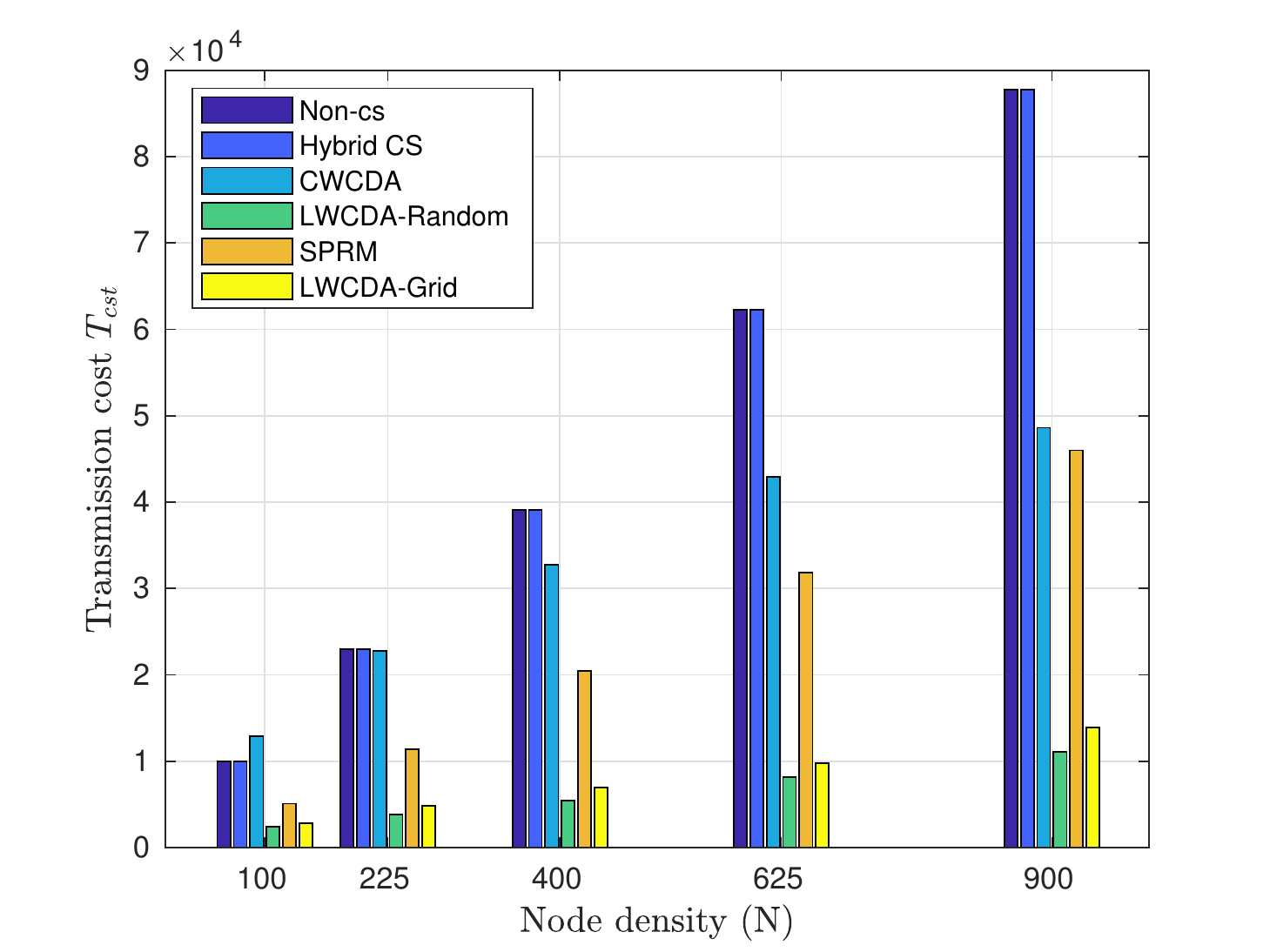}
\caption{Transmission cost comparison of the proposed LWCDA method against node density with SPRM, Hybrid CS, CWCDA and Non-CS methods at $\Gamma= 50\%$. Here, an area of $256m \times 256m$ is considered for the network deployment and number of nodes deployed ($N$)  are varied. Transmission cost of the LWCDA is significantly low compared to SPRM, Hybrid CS, CWCDA and Non-CS methods for all considered node densities.}    
\label{tcost_density}
\end{figure}

\begin{figure}
\centering
\includegraphics[scale = 0.48]{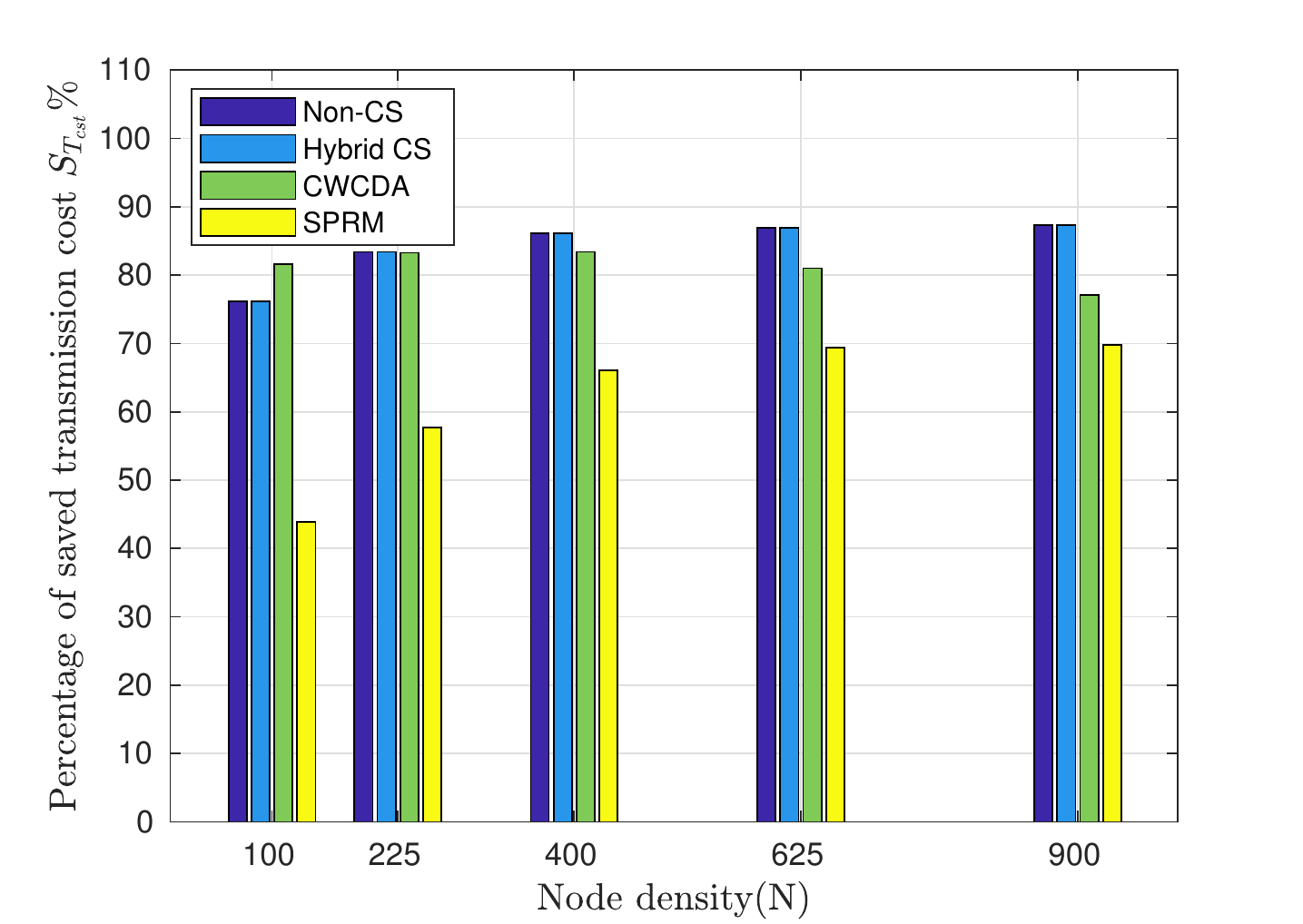}
\caption{Comparison of the percentage of saved transmission cost of the proposed LWCDA with respect to SPRM, Hybrid CS, CWCDA and Non-CS methods against node density at $\Gamma=50\%$. Here, an area of $256m \times 256m$ is considered for the network deployment and number of nodes ($N$) deployed are varied. The proposed LWCDA method offers savings in the transmission cost consistently with respect to the baseline approaches for all the considered node densities.}    
\label{tcost_density_save}
\end{figure}

\par To evaluate the effect of changing the node density on the required transmission cost for data aggregation of the proposed LWCDA, we performed an experiment where the number of nodes deployed is varied in the considered area of $256m \times 256m$. The transmission cost of data aggregation with respect to the changing in node density with $\Gamma=50\%$ compression rate is shown in Fig.~\ref{tcost_density}. From Fig.~\ref{tcost_density}, it is observed that the transmission cost increases with an increase in the node density. The interesting observation made from Fig.~\ref{tcost_density} is that the transmission cost for LWCDA is significantly low as compared to that of the traditional methods for all considered density levels in both deployment scenarios. 
Fig.~\ref{tcost_density_save} shows the percentage of savings in the transmission cost with respect to Non-CS, Hybrid CS, CWCDA and SPRM methods. Percentage of saved transmission cost $S_{T_{cst}}$ of a given method $P$ with respect to the method $Q$ is defined as,
\begin{equation}
S_{T_{cst}}\%  =\bigg(1-\frac{T_{cst}\text{ of method }P}{T_{cst}\text{ of method }Q}\bigg)\times 100.
\end{equation} 
       
From Fig.~\ref{tcost_density_save}, one can observe that the proposed method consistently offers savings in transmission cost under the considered varying node densities. We can infer that for large-scale dense networks, LWCDA algorithm can achieve significant improvements in the network lifetime compared to traditional approaches.

\begin{figure}
\centering
\includegraphics[scale = 0.48]{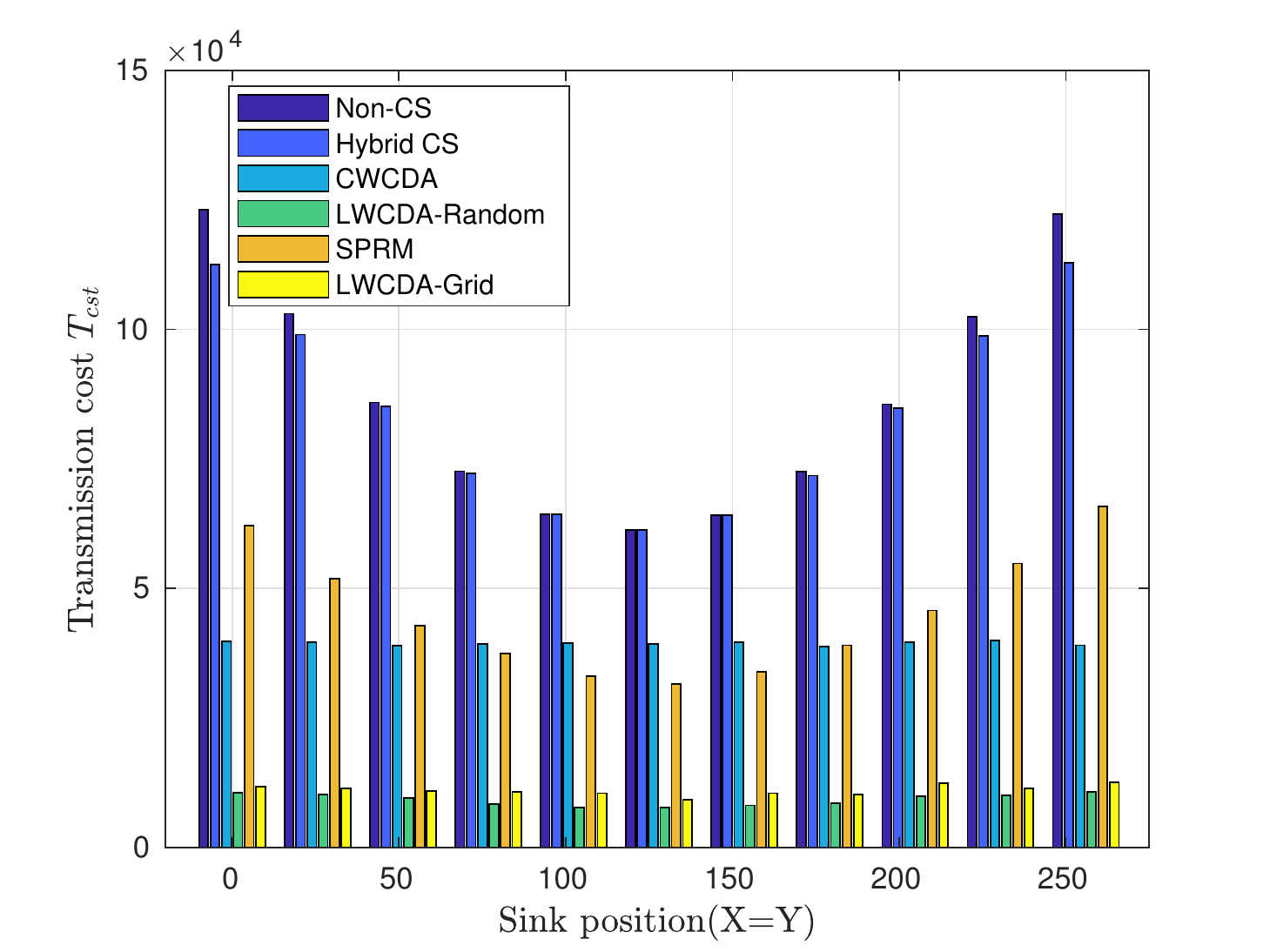}
\caption{Comparison of the transmission cost required for data aggregation from $625$ nodes deployed in an area of $256m \times 256m$ using the proposed LWCDA and SPRM, Hybrid CS, CWCDA and Non-CS methods against the sink location. The sink node location ($X$, $Y$) varies according to the line $X=Y$ where $X$, $Y \in [0,256]$. Transmission cost of the proposed LWCDA method is robust and lower compared to all baseline approaches for all considered sink locations.}    
\label{sink_pos_change}
\end{figure}

The location of the sink node affects the required transmission cost for data aggregation \cite{safa_JNCA}. To analyze the dependence of the transmission cost on the sink location for data aggregation, we considered a $625$-node network deployed (grid and random deployment) in an area of $256m \times 256m$ with varying sink locations. Fig.~\ref{sink_pos_change} compares the transmission cost of the proposed LWCDA algorithm with that of SPRM (for grid-wise deployment) and Non-CS, Hybrid CS and CWCDA (for random deployment) with respect to various sink locations. Note that the variables $X$, $Y \in [0,256]$ represent the geographic coordinates of the sink node on the considered area. The sink location ($X$, $Y$) varies on the line $X=Y$. The observation that can be made from Fig.~\ref{sink_pos_change} is that the transmission cost of baseline approaches except CWCDA strongly depends on the sink location. Transmission cost required for data aggregation with CWCDA is robust to sink location, but it requires more transmission cost compared to the proposed method across all sink locations. The considered traditional approaches (SPRM, Hybrid CS and NoN-CS) yield lower transmission cost when the sink is at the center of the considered area. In fact, if the sink is at the center of the considered area, every node can connect to the sink with the shortest distance. An interesting inference that can be made from Fig.~\ref{sink_pos_change} is that the transmission cost of the proposed LWCDA algorithm for data aggregation in both grid-wise and random deployment scenario is robust to the sink location and is much lower compared to that of the traditional methods for all the considered sink locations. This can be justified by noting that the proposed LWCDA algorithm aggregates data through clustering where required transmission cost is independent of the sink location. In addition, to aggregate measurements from randomly distributed CHs, which are connected through MST along with the sink node, incur almost same transmission cost irrespective of the sink location.

The data aggregation algorithms proposed in the literature such as \cite{cso}-\cite{rndmwlk} do not discuss hardware implementation details. The implementation procedure and assumptions considered for software simulations differ when it comes to real-time hardware implementation. For the completeness of the proposed algorithm and analysis, we describe an optimal way of implementation which shows the efficacy of the proposed method in a real-time scenario.

\section{Real-time Implementation}
The in-house IITH Motes \cite{mote} are used for implementing the proposed data aggregation algorithm (LWCDA). The IITH Mote is a ZigBee system-on-chip combining a 2.4 GHz IEEE 802.15.4 radio transceiver with a 8 MHz, 8-bit processor having 128 kB of flash memory and 8 kB of RAM. TinyOS \cite{tinyos} is used to program the proposed LWCDA algorithm on the nodes. Based on the required compression rate $\Gamma$, the threshold value $T_{hr}$ will be decided. From the selected $M$ CHs, the probability of the $i^{th}$ node becoming a CH is $P_{CH}=\frac{M}{N}$ as discussed in Section III. Let $T_{u_{i}}$ denote the generated uniform random value at the $i^{th}$ node, i.e., $T_{u_{i}} \in U\sim [0,1]$. If $T_{u_{i}}\leq T_{hr}$ then the $i^{th}$ node becomes a CH. CH probability can be rewritten as $P_{CH}= Pr\{T_{u_{i}}\leq T_{hr}\}= T_{hr}$ where $i \in [1,N]$. For example, if the threshold is considered to be $T_{hr} = 0.3$ then on an average $30\%$ of the nodes become CHs ($P_{CH}=0.3$) and $\Gamma = 70\%$ compression can be achieved. The sink node broadcasts a starting packet with the specified threshold $T_{hr}$ value. Each node in the network broadcasts this packet once so that the threshold value reaches every other node in the network. The nodes calculate Received Signal Strength Indication (RSSI) values from the received packets and stores them in a table. It is important to note that each node will have RSSI values of all the other nodes that are in its radio range (communication range). Using the created RSSI table, the nodes, which are selected as leaf nodes, connect to nearer CHs and CHs form MST. 
\begin{algorithm}[t]
    \caption{Pseudo code for the data aggregation algorithm at node level}\label{alg1}
    \begin{algorithmic}[1]
    \REQUIRE $T_{hr}$  
    \STATE Data collection round $r=0$ 
    \STATE Generate uniform random value $T_{u_{i}}\in U\sim [0,1]$ ($i$ refers node number)
    \IF{$T_{u_{i}}\leq T_{hr}$} 
      \STATE $Type$ = $CH$
    \ELSE
      \STATE $Type$ = $Leaf node$
    \ENDIF
    \WHILE{$r\geq 0$}
    \IF{$Type==CH$}
    \STATE $r=r+1$
    \IF{$r==1$}
    \STATE Broadcast CH packet
    \STATE Generate uniform random value $R_{i} \in U\sim [0,1]$
      \IF{$R_{i}\leq 0.5$}
    \STATE $\alpha_{i}= -1$ 
      \ELSE 
    \STATE $\alpha_{i} = 1$
      \ENDIF
    \STATE Discover the next hop destination node $CH_{dest}$: CH node or the leaf node in MST towards the sink  
     \ENDIF 
    \STATE Measure data sample $x_{i}$ 
    \STATE Compute: $\alpha_{i}x_{i}$
    \STATE Receive data packets from all the leaf nodes and descendant CHs
    \STATE Compute: $\sum_{i\in c_{j}}\alpha_{i}x_{i}$ 
    \STATE Send CH data packet to $CH_{dest}$ using pack and forward method
    \ELSE  
      \STATE $r=r+1$
      \IF{$r==1$}
      \STATE Find $RSSI_{h}=$ $\max \limits_{h} \{ RSSI $of CHs which are in the radio range$\}$
      \STATE $Leaf_{dest}$ = $CH_{h}$
      \IF{$Leaf_{dest}==NULL$}
      \STATE Discover the next hop destination node $Leaf_{dest}$ = neighboring leaf node in the shortest path towards nearer CH 
      \ENDIF 
      \STATE Generate uniform random value $R_{i} \in U\sim [0,1]$
      \IF{$R_{i}\leq 0.5$}
      \STATE $\alpha_{i}= -1$
      \ELSE 
      \STATE $\alpha_{i} = 1$
      \ENDIF
      \ENDIF
      \STATE Measure data sample $x_{i}$
      \STATE Compute: $\alpha_{i}x_{i}$
      \STATE Send the data packet to $Leaf_{dest}$
    \ENDIF
    \ENDWHILE
\end{algorithmic}
\end{algorithm}

\par As the sink node requires the knowledge of $\Phi$, i.e., $\{\alpha_{i}\}$ values and respective indices $\Delta_{j}$, where $i\in c_{j}$ and $j\in[1,M]$, in the initial phase (i.e., first cycle of data aggregation), CH sends $\{\alpha_{i}\}$, $\Delta_{j}$ to the sink along with the final measurement $\sum_{i\in c_{j}}\alpha_{i}x_{i}$. This is a small overhead as $\alpha$ and the respective node index (node address) together can take a maximum of three octets when short address mode is considered. By the end of the initial phase, all the nodes register their respective destination node addresses. In data sensing phase (i.e., from second data aggregation cycle on-wards), in each cycle, all the leaf nodes compute their measurements and send them to their respective destined CHs. Further, each CH computes the final measurement and forwards it to the sink. Pseudo code of the node level implemented algorithm is described in Algorithm \ref{alg1}.

\begin{figure}[ht]
\centering
\includegraphics[scale = 0.4]{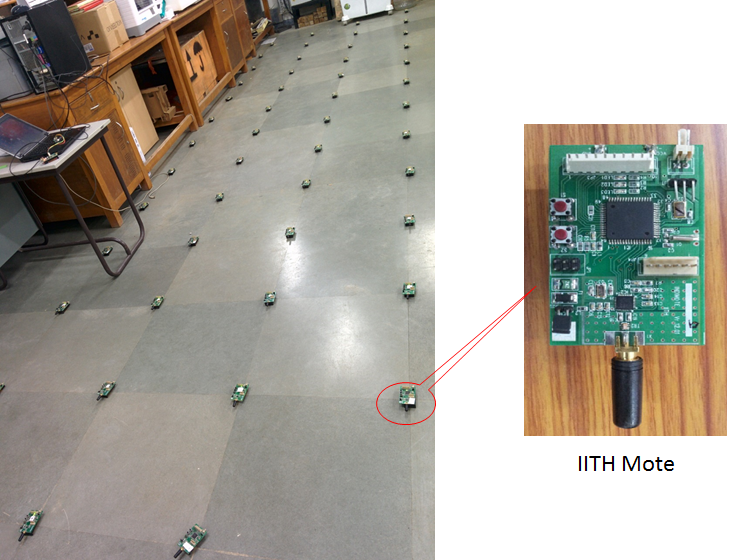}
\caption{Experimental setup with $N=50$ nodes deployed in an area of $321.44$ $ft^{2}$.}
\label{deployment}
\end{figure}
 
\begin{figure}[!t]
\centering
\includegraphics[scale = 0.5]{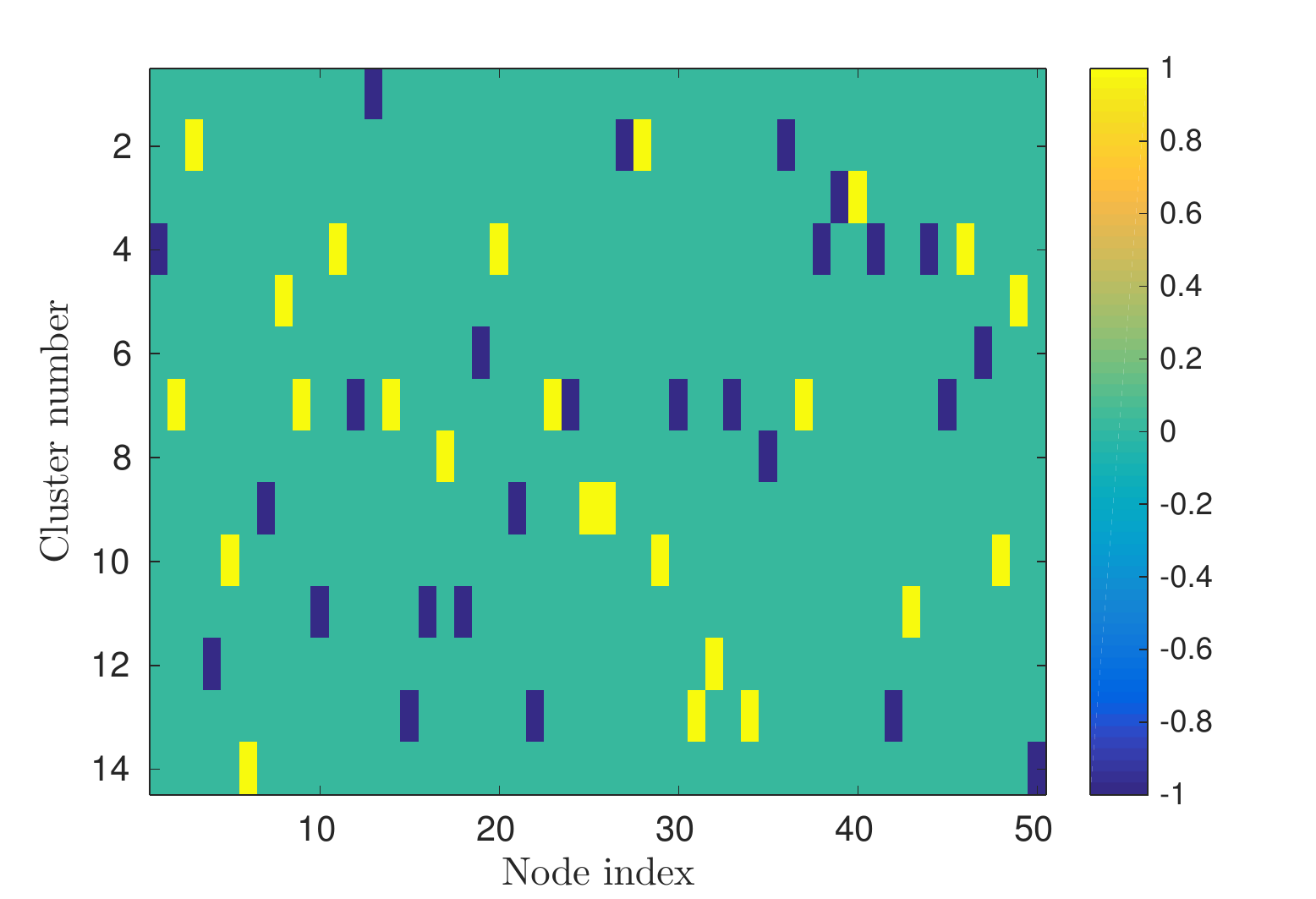}
\caption{Measurement matrix $\Phi_{14 \times 50}$ constructed from the real field deployment with $N=50$ nodes and $\Gamma =70\%$.}
\label{Mtc}
\end{figure}

\begin{figure}[!t]
\centering
\includegraphics[scale = 0.4]{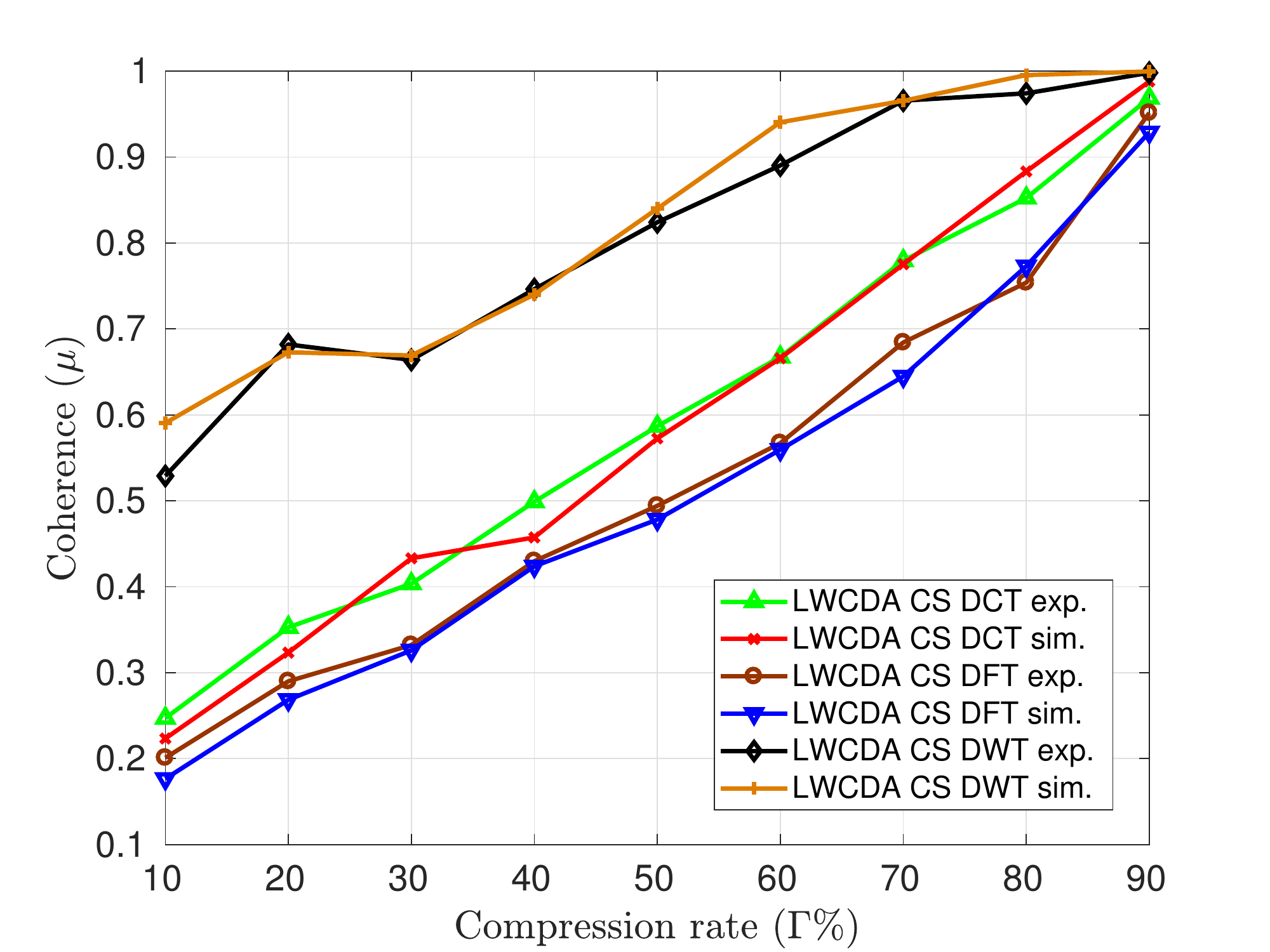}
\caption{Comparison of the mutual coherence $\mu$ for the CS matrix $A=\Phi\Psi$ where $\Phi$ is constructed from the real-time deployment and simulations against different compression rates $\Gamma$, with $N=50$ nodes and for different bases DFT, DCT and DWT. Mutual coherence curves from simulation and real-time deployment are very close and demonstrate the efficacy of the proposed method.}
\label{exprmnt}
\end{figure}

The proposed LWCDA algorithm is independent of the deployment scenario. As an example to verify the implementation efficacy of the LWCDA, we deployed $50$ nodes grid-wise in an area of $321.44$ $ft^{2}$ as shown in Fig.~\ref{deployment}. The sink node is connected to a PC that collects measured data from all the nodes in the network. For illustration, we considered a threshold $T_{hr}=P_{CH}=0.3$ and obtained $14$ CHs among the deployed $50$ nodes in a particular realization of the experiment, while the remaining nodes are connected to their respective CHs. Accordingly, $14$ clusters were formed, and thus the rows of the measurement matrix $\Phi_{14 \times 50}$ were generated. The resultant measurement matrix $\Phi_{14 \times 50}$ is shown in Fig.~\ref{Mtc}. To evaluate the efficacy of the proposed method, we considered coherence as the metric. We repeated the above experiment for a range of threshold values $T_{hr}=0.1:0.1:0.9$, and in each case of $T_{hr}$, the measurement matrix $\Phi$ was constructed. To compute the coherence of $\Phi$ against all the compression rates, we obtained $\Phi$ for $10$ realizations and for each $T_{hr}$. Each realization gives one mutual coherence value $\mu$ for a pair of $\Phi$ and $\Psi$. We then averaged $\mu$ over $10$ realizations for each $T_{hr}$. To compare with the real deployment, we simulate a similar scenario in software. Average coherence values of the matrix $A$ designed from both the experiment (exp.) as well as the simulation (sim.) are plotted in Fig.~\ref{exprmnt}. Fig.~\ref{exprmnt} illustrates that the coherence values of the matrix $A$ where the proposed measurement matrix is constructed from the experiment as well as the simulation with DFT, DCT and DWT bases are in excellent agreement. These results show efficacy of the proposed method in a real-time implementation. It justifies our claim that the proposed method does not require any extra computational overhead (such as the generation of the individual columns of the matrix $\Phi$, storage of $\Phi$ etc.). Hence, the proposed method can be implemented on low end commercial off-the-shelf IoT nodes.

\section{Conclusion}
In this paper, we proposed a CS based data aggregation method for IoT networks which is both light weight (possessing low complexity) and energy efficient. In the proposed data aggregation algorithm, to minimize the transmission cost, data is aggregated from non-overlapping clusters where each node can contribute to one measurement. Hence, the columns of the measurement matrix constructed from the proposed algorithm are coherent and recovery is not possible for the data which is sparse in the canonical basis (Identity matrix). However, we showed that the measurement matrix when combined with the popular bases (DFT, DCT, DWT, Laplacian and DiWT) can guarantee the recovery of data with high fidelity. 
 
Unlike conventional methods, in the proposed data aggregation method the node-level complexity is independent of the network size and data sparsity. The comparison of the transmission cost concludes that the proposed method is energy efficient and can aid in extending the network lifetime by achieving minimal transmission cost. Hardware implementation demonstrated the efficacy of the proposed algorithm in a real-time implementation. Further, through the analysis of the measurement matrix combined with the popular bases, we found that our data aggregation method using the DFT basis yields a better reconstruction quality than other bases. However, it is still unknown whether there exists a relation between the measurement matrix and the DFT basis. We hope to provide a deeper insight in our future investigations and present theoretical guarantees. We observed that there is a slight variation in the performance obtained through DFT and DCT bases. In future, we will pursue a thorough analysis of this discrepancy in the performance variation and study the behavior of energy consumption of the proposed method in the presence of interference.

\end{document}